\documentclass[aps,prd,nofootinbib,showkeys,final,floatfix,superscriptaddress]{revtex4-1}
\pdfoutput=1

\usepackage{amssymb}
\usepackage{amsmath}
\usepackage{amsfonts}
\usepackage{graphicx}
\usepackage{xcolor}
\usepackage{xspace}
\usepackage{ wasysym }

\usepackage{float}
\usepackage{subfig}
\usepackage{mathrsfs}
\usepackage{slashed}

\def\gsim{\raise0.3ex\hbox{$\;>$\kern-0.75em\raise-1.1ex\hbox{$\sim\;$}}}
\def\lsim{\raise0.3ex\hbox{$\;<$\kern-0.75em\raise-1.1ex\hbox{$\sim\;$}}}

\newcommand{\ba}[1]{\begin{eqnarray} \label{(#1)}}
\newcommand{\ea}{\end{eqnarray}}

\newcommand{\AddrAHEP}{
  AHEP Group, Instituto de F\'{\i}sica Corpuscular --
  CSIC/Universitat de Val{\`e}ncia \\
  Edificio de Institutos de Paterna, Apartado 22085,
  E--46071 Val{\`e}ncia, Spain}

\newcommand{\AddrUSerena}{
  Departamento de F\' isica, Facultad de Ciencias, Universidad de La Serena,\\
  Avenida Cisternas 1200, La Serena, Chile. 
}

\newcommand{\AddrPrague}{ Institute of Particle and Nuclear Physics \\
  Faculty  of  Mathematics  and  Physics,  Charles  University,\\
  V Hole\v{s}ovi\v{c}k\'ach 2, 18000 Prague 8, Czech Republic}

\newcommand{\AddrUFSM}{
  Universidad T\'ecnica Federico Santa Mar\'\i a, \\ 
  Casilla 110-V, Valpara\'\i so,  Chile}

\newcommand{\AddrCCTVal}{
  Centro-Cient\'\i fico-Tecnol\'{o}gico de Valpara\'\i so, \\ 
  Casilla 110-V, Valpara\'\i so,  Chile}

\usepackage[linktocpage,colorlinks=true,urlcolor=blue!80!red,linkcolor=blue,citecolor=red]{hyperref}

\begin{document}

\preprint{IFIC/18-XX}

\title{High-dimensional neutrino masses}

\author{Gaetana Anamiati}
\email{anamiati@ific.uv.es}
\affiliation{\AddrAHEP}

\author{Oscar Castillo-Felisola}
\email{o.castillo.felisola@gmail.com}
\affiliation{\AddrUFSM}
\affiliation{\AddrCCTVal}

\author{Renato M. Fonseca}
\email{fonseca@ipnp.mff.cuni.cz}
\affiliation{\AddrPrague}

\author{J.~C. Helo}
\email{jchelo@userena.cl}
\affiliation{\AddrUSerena}
\affiliation{\AddrCCTVal}

\author{M. Hirsch}
\email{mahirsch@ific.uv.es}
\affiliation{\AddrAHEP}



\begin{abstract}
  
  For Majorana neutrino masses the lowest dimensional operator
  possible is the Weinberg operator at $d=5$. Here we discuss the
  possibility that neutrino masses originate from higher dimensional
  operators. Specifically, we consider all tree-level decompositions
  of the $d=9$, $d=11$ and $d=13$ neutrino mass operators. With
  renormalizable interactions only, we find 18 topologies and 66
  diagrams for $d=9$, and 92 topologies plus 504 diagrams at the
  $d=11$ level. At $d=13$ there are already 576 topologies and 4199
  diagrams. However, among all these there are only very few genuine
  neutrino mass models: At $d=(9,11,13)$ we find only (2,2,2) genuine
  diagrams and a total of (2,2,6) models. Here, a model is considered
  genuine at level $d$ if it automatically forbids lower order
  neutrino masses {\em without} the use of additional symmetries. We
  also briefly discuss how neutrino masses and angles can be easily
  fitted in these high-dimensional models.

\end{abstract}

\maketitle

%

\section{Introduction}
\label{sect:intro}

The Weinberg operator is the lowest dimensional non-renormalizable
operator that one can write down with only standard model (SM)
fields~\cite{Weinberg:1979sa}. It violates lepton number by two units
and thus, once the electro-weak symmetry is broken, Majorana neutrino
masses are generated.  The observed smallness of the neutrino masses
is then usually attributed to the large value of the scale of lepton
number violation (LNV), typically $\Lambda \sim (10^{14}-10^{15})$
GeV. This is the essence of the seesaw
mechanism~\cite{Minkowski:1977sc,Yanagida:1979as,GellMann:1980vs,Mohapatra:1979ia,Magg:1980ut,Schechter:1980gr,Wetterich:1981bx,Lazarides:1980nt,Mohapatra:1980yp,Cheng:1980qt,Foot:1988aq}. While
simple and elegant, the large mass scale involved in this argument
makes direct tests of the classical seesaw impossible.

There exist, however, many possibilities to explain the smallness of
the observed neutrino masses with lower LNV scales. For Majorana
neutrinos one can write in general~\cite{Bonnet:2012kz}
\begin{equation}
  \label{eq:mnugen}
  m_{\nu} \propto 
  \epsilon 
  \cdot
  \left( \frac{1}{16 \pi^{2}} \right)^{n}
  \cdot 
  \left(\frac{v}{\Lambda}\right)^{d-5}
  \cdot
  \frac{v^2}{\Lambda} .
\end{equation}
Here, $v$ stands for the standard model vacuum expectation value
(vev), $d$ is the dimension of the operator, $n$ stands for the number
of loops at which neutrino masses are generated. $\epsilon$ expresses
symbolically the additional suppression of lepton number violation
that might arise in particular constructions, such as for example the
inverse seesaw mechanism~\cite{Mohapatra:1986bd}. Finally, in
addition, small Yukawa or scalar couplings, not shown explicitly in
Eq.~\eqref{eq:mnugen}, could lead to smaller than expected neutrino
masses.

Equation~\eqref{eq:mnugen} can be used to estimate the typical scale
$\Lambda$, for which the observed neutrino masses could be explained
for a given $d$ and $n$. Fig.~\ref{fig:scales} illustrates this
estimate.  Here, ${\cal O}_5$ at tree-level corresponds to the
classical seesaw mechanism. Note that for ${\cal O}_5$ at tree-level
(1-loop level) Yukawa couplings of order ${\cal O}(10^{-6})$ (${\cal
  O}(10^{-3})$) would be needed to obtain a scale as low as $\Lambda
\simeq 1$ TeV.  In this figure we also show the estimated reach for
three colliders. The LEP line reflects that no electrically charged
particle coupled to SM fermions with masses below roughly 100 GeV can
exist, after the negative searches performed at the LEP
collider~\cite{Patrignani:2016xqp}. The horizontal grey band indicates
a very rough estimate of the reach of the LHC: The lower edge of the
band is a more conservative estimate (pair production of charged
particles), while the upper edge is roughly the reach of the LHC for
particles produced in s-channel diagrams and/or with colour.  For
$d=9$ and larger one expects that LHC experiments will cover an
important part of the available parameter space of these models. We
also show as a dashed line a rough estimate of the reach of a
hypothetical $\sqrt{s}=100$ TeV collider, here called FCC. Thus,
neutrino mass models generated at $d=9$ and higher should be testable
in the near future.  This simple argument forms the main motivation
for our current paper.
\begin{center}
  \begin{figure}
    \begin{centering}
      \includegraphics[scale=0.6]{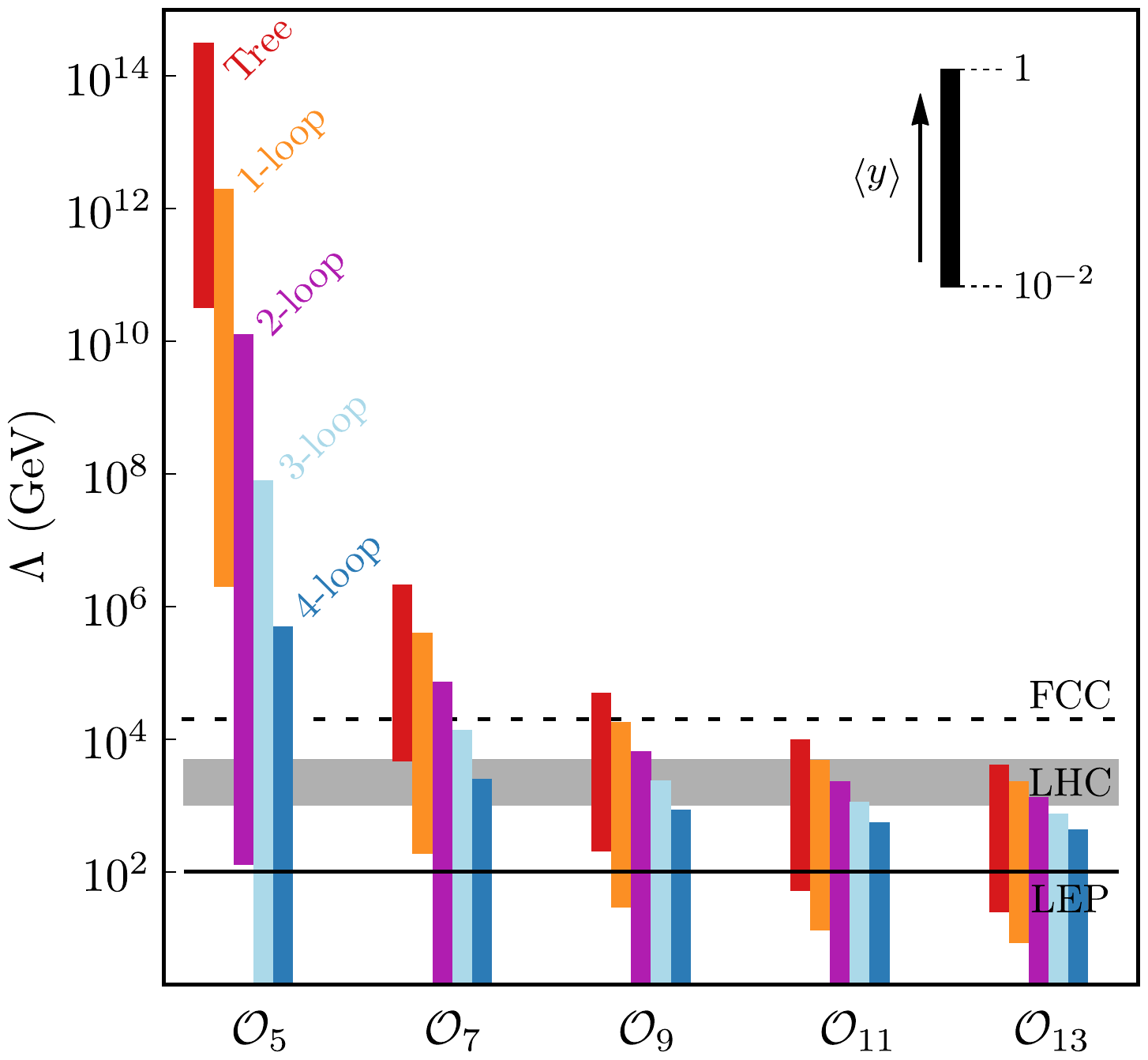}
    \end{centering}
    \caption{\label{fig:scales} The typical energy scales ($\Lambda$)
      for which a neutrino mass model with a given dimension and
      number of loops $(d,n)$ can explain correctly the observed
      sub-eV neutrino masses. Operators start at $d=5$, corresponding
      to the Weinberg operator.  Energy ranges have been estimated
      using average couplings $\left\langle y\right\rangle$ in the
      range of $[0.01,1]$.  }
  \end{figure}\par
\end{center}

Here, we will study high-dimensional tree-level diagrams for Majorana
neutrino masses. We will treat systematically all possible topologies
for the deconstruction of the $d=9$, $d=11$ and $d=13$ operators. We
will identify all the ``genuine'' diagrams, which for us are those
diagrams that can give the leading contribution to the neutrino mass
matrix, without the use of extra (discrete or flavour) symmetries.
We will discuss this requirement in more detail in section
\ref{subsect:gen}.  Despite the large number of possible topologies,
for $d=9$ and $d=11$ surprisingly only 4 models survive: 2 at $d=9$
and 2 at $d=11$.  For $d=13$ we have found a total of 2
genuine diagrams and 6 models that can realize them.

Before presenting our analysis, let us briefly mention that, of
course, many authors have studied neutrino mass models beyond the
simplest tree-level seesaw, for a recent review see for
example~\cite{Cai:2017jrq}.  The Zee model~\cite{Zee:1980ai}, or the
Zee-Babu model~\cite{Cheng:1980qt,Zee:1985id,Babu:1988ki} are early
examples of 1-loop and 2-loop realizations of the Weinberg operator. A
systematic analysis of possible neutrino mass models at $d=5$ and
1-loop can be found in Ref.~\cite{Bonnet:2012kz}, for a general
analysis of $d=5$ models at 2-loop see Ref.~\cite{Sierra:2014rxa}. For
the 3-loop case, there exist some well-known models in the
literature~\cite{Krauss:2002px,Gustafsson:2012vj}; a complete study of
3-loop neutrino masses at $d=5$ can be found in
Ref.~\cite{Cepedello:2018aaa}. Neutrino masses at $d=7$ level have
also been studied. A systematic analysis at tree-level was done in
Ref.~\cite{Bonnet:2009ej}. At $d=7$ tree-level there is only one
genuine (in our sense) tree-level neutrino mass model, which was first
discussed in Ref.~\cite{Babu:2009aq}; we will call it the BNT model
below. A general analysis of $d=7$ neutrino masses at 1-loop order was
recently presented in Ref.~\cite{Cepedello:2017eqf,Cepedello:2017lyo}.

Then there are also some papers on $d=9$ (and higher) neutrino mass
models, see
Refs.~\cite{Picek:2009is,Kumericki:2012bh,Liao:2010cc,McDonald:2013kca,McDonald:2013hsa,Nomura:2017abu}. We
will come back to these papers briefly in Sec.~\ref{sect:prelim},
where we discuss the main differences between their results and our
present work. We mention in passing also the model presented in
Ref.~\cite{Nomura:2016jnl}, which uses a scalar septet to construct a
model giving $d=13$ neutrino masses at 1-loop. Note, however, that
this model is not genuine in our sense, since it uses a $Z_2$ symmetry
to eliminate the $d=5$ seesaw contribution.

The rest of this paper is organized as follows. In
Sec.~\ref{sect:prelim} we will set up our notation and briefly discuss
neutrino mass generation at lower dimensions. This is necessary to
clearly define what we mean by ``genuine''
models. Section~\ref{sect:class} then contains the central piece of
our work. We explain our methods, discuss topologies and list and
briefly discuss the genuine models.  In Sec.~\ref{sect:cncl} we give a
short conclusion. In the appendix we discuss how experimental data on
neutrino masses and mixing can be easily fitted with these
high-dimensional models.

\section{Preliminaries\label{sect:prelim}}

In this section we briefly go over some basic facts about $d=5$ and $d=7$
neutrino masses. This will be useful later, when we discuss genuine
higher dimensional models, since those models can give the dominant
contribution to the neutrino mass matrix only if $d=5$ and $d=7$
contributions are absent. We will use the following notation.
A $SU(2)_L$ multiplet with hypercharge $Y$ is denoted as ${\bf R}_Y$,
to which we add the superscript $F$ or $S$ for fermion or scalar,
respectively. Thus, for example ${\bf 5}_0^F$ is a hypercharge-less
fermionic quintuplet.

\subsection{Tree level $d=5$ and $d=7$}

The $d=5$ Weinberg operator can be generated at tree-level in exactly
three different ways~\cite{Ma:1998dn}. In the literature these are
known as seesaw type-I, type-II and type-III. Type-I is the standard
contribution due to right-handed neutrino $\nu_R$ (or ${\bf 1}_0^F$ in
our notation).  The Majorana mass term for $\nu_R$ is the origin of
lepton number violation. Type-III seesaw replaces ${\bf 1}_0^F$ by
${\bf 3}_0^F$~\cite{Foot:1988aq}, which is a field usually denoted as
$\Sigma$ in the literature. Finally, for type-II seesaw one introduces
${\bf 3}_1^S \equiv \Delta$. In this latter case, the presence of both
the Yukawa coupling $L\Delta L$ and the scalar coupling
$H\Delta^{\dagger}H$, leads to lepton number violation.

\begin{figure}
  \centering
  \includegraphics[scale=0.6]{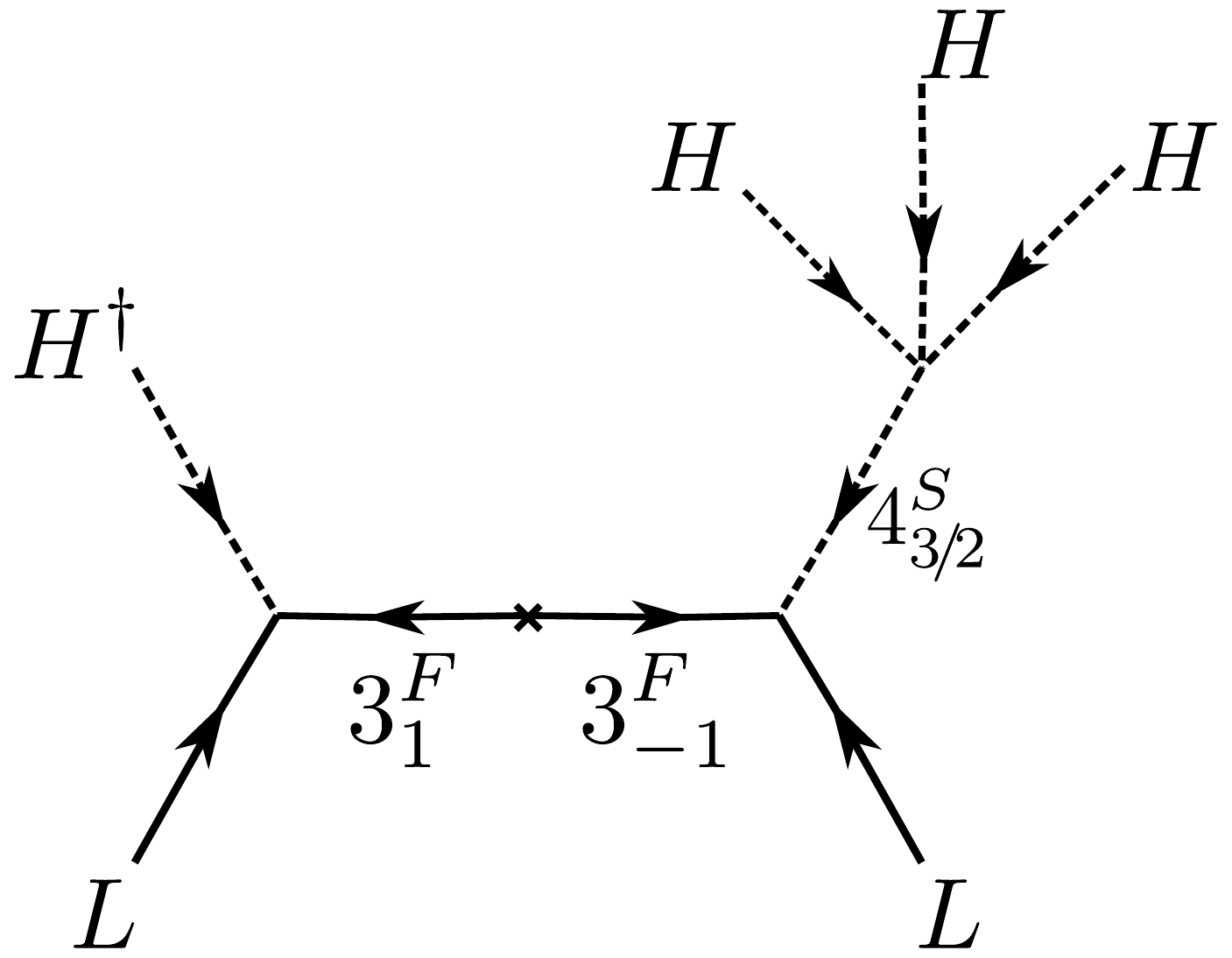}
  \caption{The genuine tree-level diagram for
    $d=7$~\cite{Babu:2009aq}. At least two beyond-the-SM particles are
    needed for higher dimensional operators.
  }
  \label{fig:d7tree}
\end{figure}

At $d=7$ one already finds five different
topologies~\cite{Bonnet:2009ej}.  However, one of these can not lead
to any renormalizable neutrino mass model, while for three more
topologies the diagrams always contain necessarily one of the $d=5$
seesaw mediators. The only diagram for which the $d=5$ tree-level
seesaw is absent without the need of additional symmetries was first
discussed in Ref.~\cite{Babu:2009aq}. This model contains two new
particles, ${\bf 3}_1^F$ and ${\bf 4}_{3/2}^S$, as shown in
Fig.~\ref{fig:d7tree}.

A few comments might be in order at this point. As
Fig.~\ref{fig:d7tree} indicates, two Weyl fermions are actually
needed to generate the diagram: ${\bf 3}_1^F$ and its vector partner
${\bf 3}_{-1}^F$.  Without the mass term $m_{{\bf 3}_1^F} {\bf
  3}_1^F{\bf 3}_{-1}^F$ there would not be any source of lepton number
violation in the model and, thus, no Majorana masses for the light,
active neutrinos could be generated.  We have therefore shown this
mass insertion explicitly in Fig.~\ref{fig:d7tree}. In many of the
diagrams in the rest of this paper, on the other hand, for a
more compact presentation, we do not explicitly show the vector
partners. However, we stress that in all of our tree-level models all
exotic fermions must necessarily be of vector-type or Majorana
fermions. Also, while at $d=5$ one new particle is sufficient for each
of the three seesaws, at $d=7$ we already need two different fields
(three if one counts the vector fermion as two distinct Weyl fermions)
for a genuine model.

\subsection{1-loop $d=5$ and $d=7$ diagrams}

The authors of Ref.~\cite{Bonnet:2012kz} systematically analyzed all
1-loop $d=5$ topologies.  In total, there are 6 topologies, but only
two of them (called T-1 and T-3) can yield genuine models in our
sense. These lead to four different diagrams, shown in
Fig.~\ref{fig:loopd5}.  T-1-ii corresponds to the diagram of the
well-known Zee model~\cite{Zee:1980ai}, an example for T-3 is the
scotogenic model~\cite{Ma:2006km}. Note also that in all 1-loop
diagrams at least two beyond-the-SM fields are needed.

\begin{center}
  \begin{figure}[tbph]
    \begin{centering}
      \includegraphics[width=0.3\textwidth]{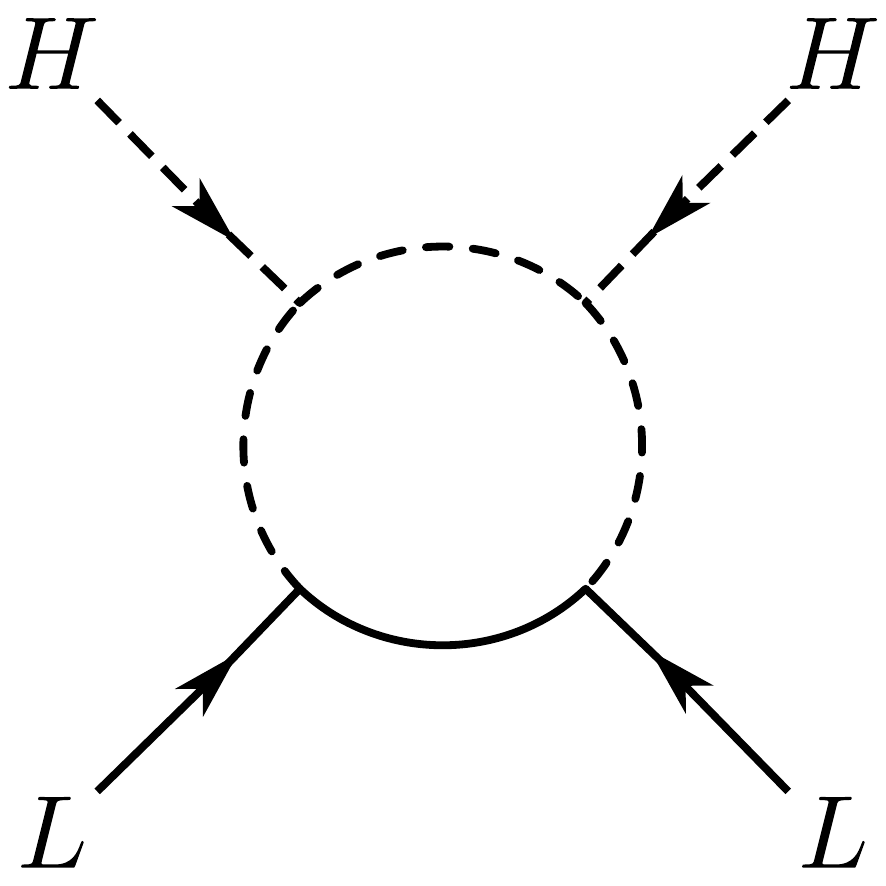}
      \includegraphics[width=0.3\textwidth]{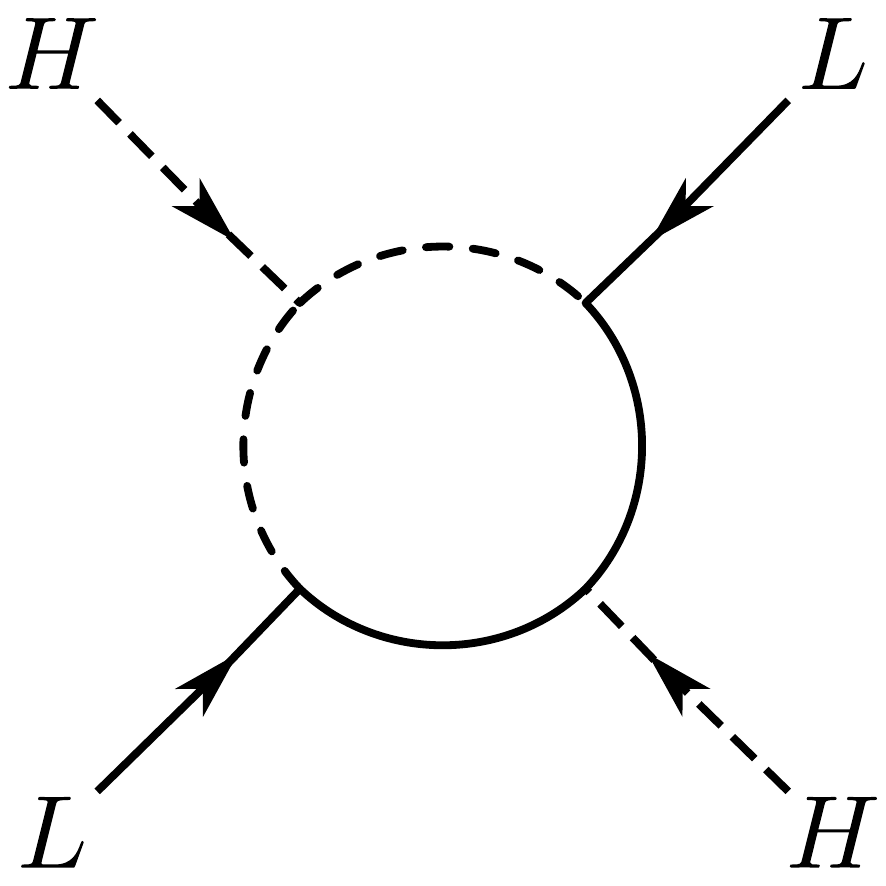}
      \vskip6mm
      \includegraphics[width=0.3\textwidth]{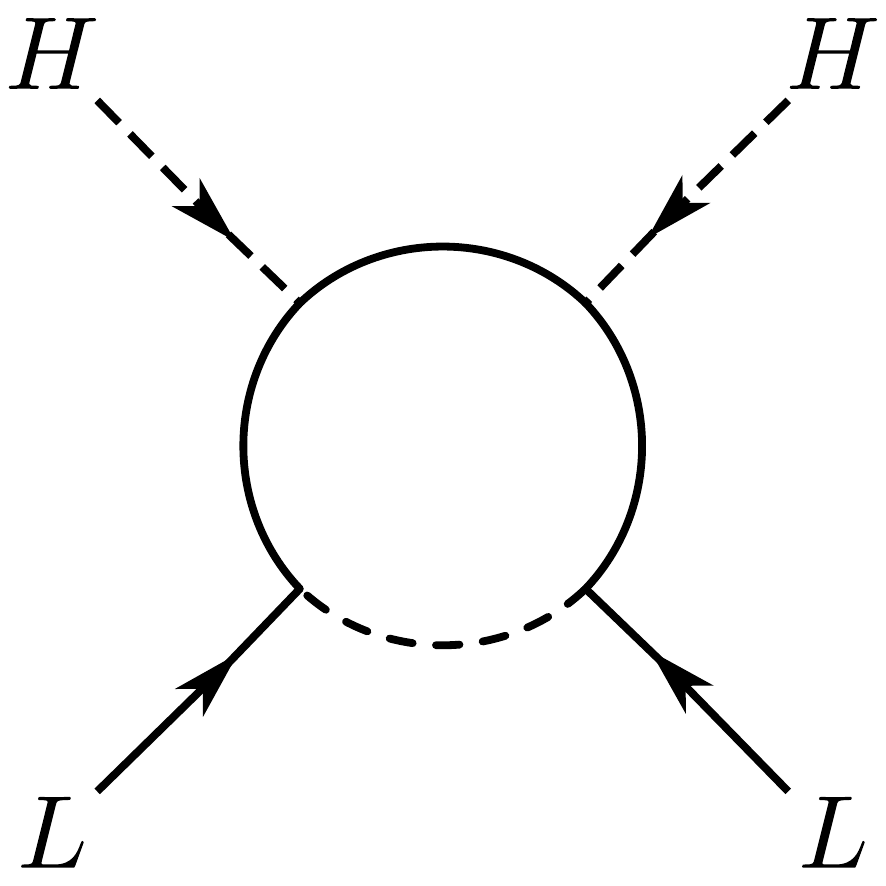}
      \includegraphics[width=0.3\textwidth]{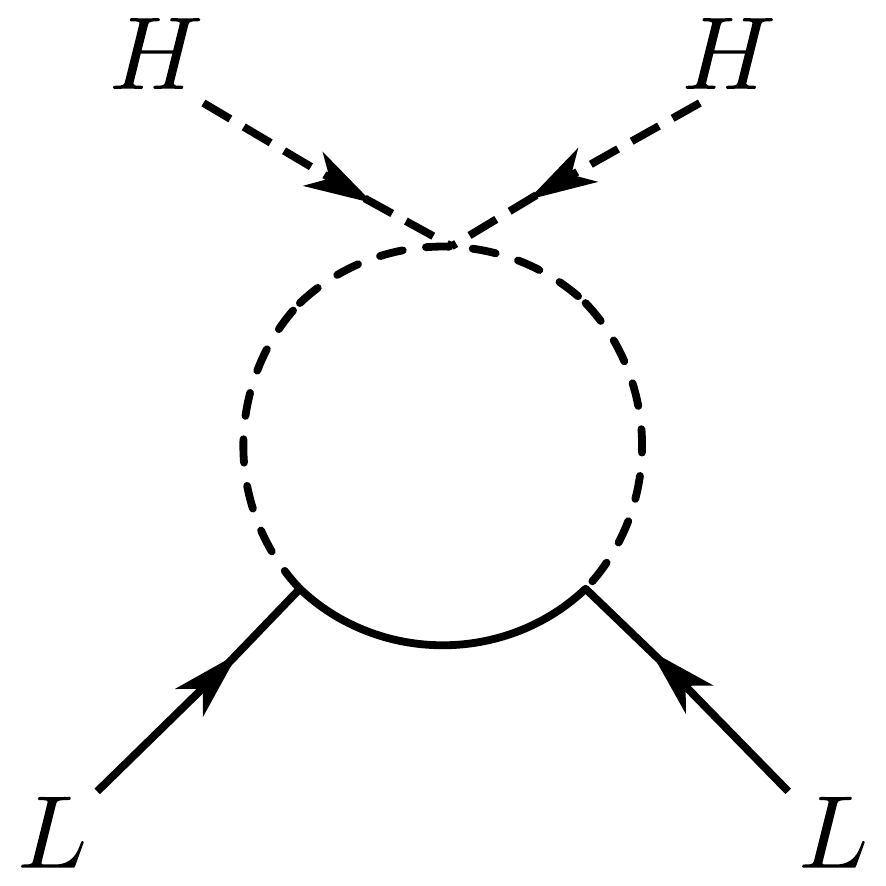}
    \end{centering}
    \protect\caption{\label{fig:loopd5} The four genuine 1-loop $d=5$
      neutrino mass diagrams~\cite{Bonnet:2012kz}. In the top panel we show the T-1-i diagram on the left and the T-1-ii diagram on the right. In the bottom panel, the T-1-iii diagram on the left and the T-3 diagram on the right.}
  \end{figure}  \par
\end{center}

In contrast to tree-level diagrams, discussed above, for 1-loop
diagrams the representation and hypercharges of the internal particles
are not uniquely fixed. Since both $L$ and $H$ are $SU(2)_L$ doublets,
the two internal particles they meet in a trilinear vertex must
transform as $({\bf N})_Y^{S/F}$ and $({\bf N\!+\!1})_{Y \pm 1/2}^{S/F}$, for
some unconstrained ${\bf N}$ and $Y$.\footnote{Although less important for
  us, we mention that differently from the tree-level realizations,
  internal particles in the loops can also be coloured. In analogy to
  what happens for the $SU(2)$ quantum numbers, since $L$ and $H$ are
  both colorless, the two internal particles which they meet in a
  trilinear vertex must transform as ${\bf R}$ and
  $\mathbf{\bar{R}}$ under $SU(3)_C$, with $\mathbf{R}$ being
  arbitrary.} This leads to a series of possible models at 1-loop, if
one allows for larger $SU(2)$ representations and hypercharges.

\begin{figure}
  \centering
  \includegraphics[width=0.45\textwidth]{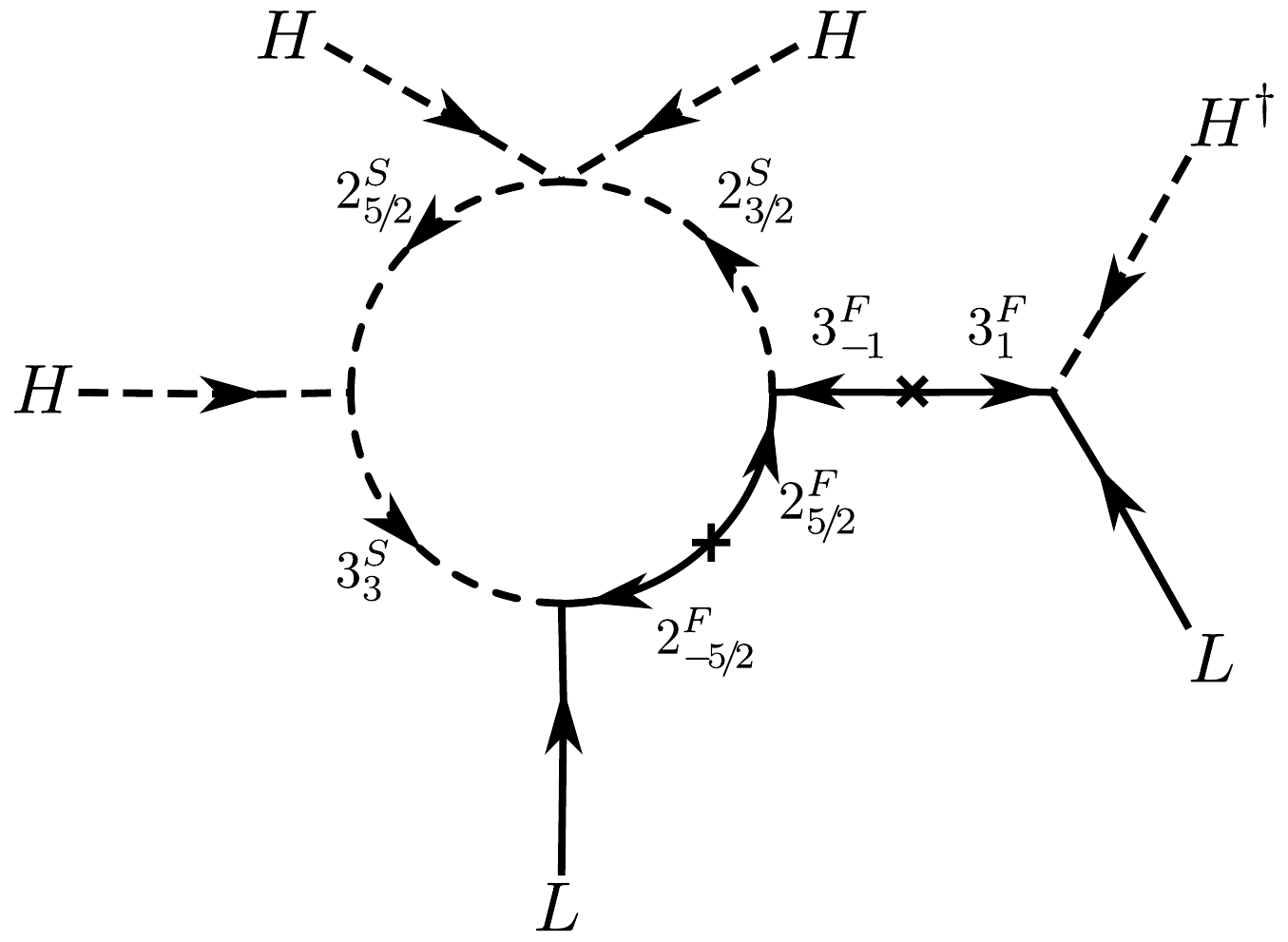}\hskip10mm
  \includegraphics[width=0.45\textwidth]{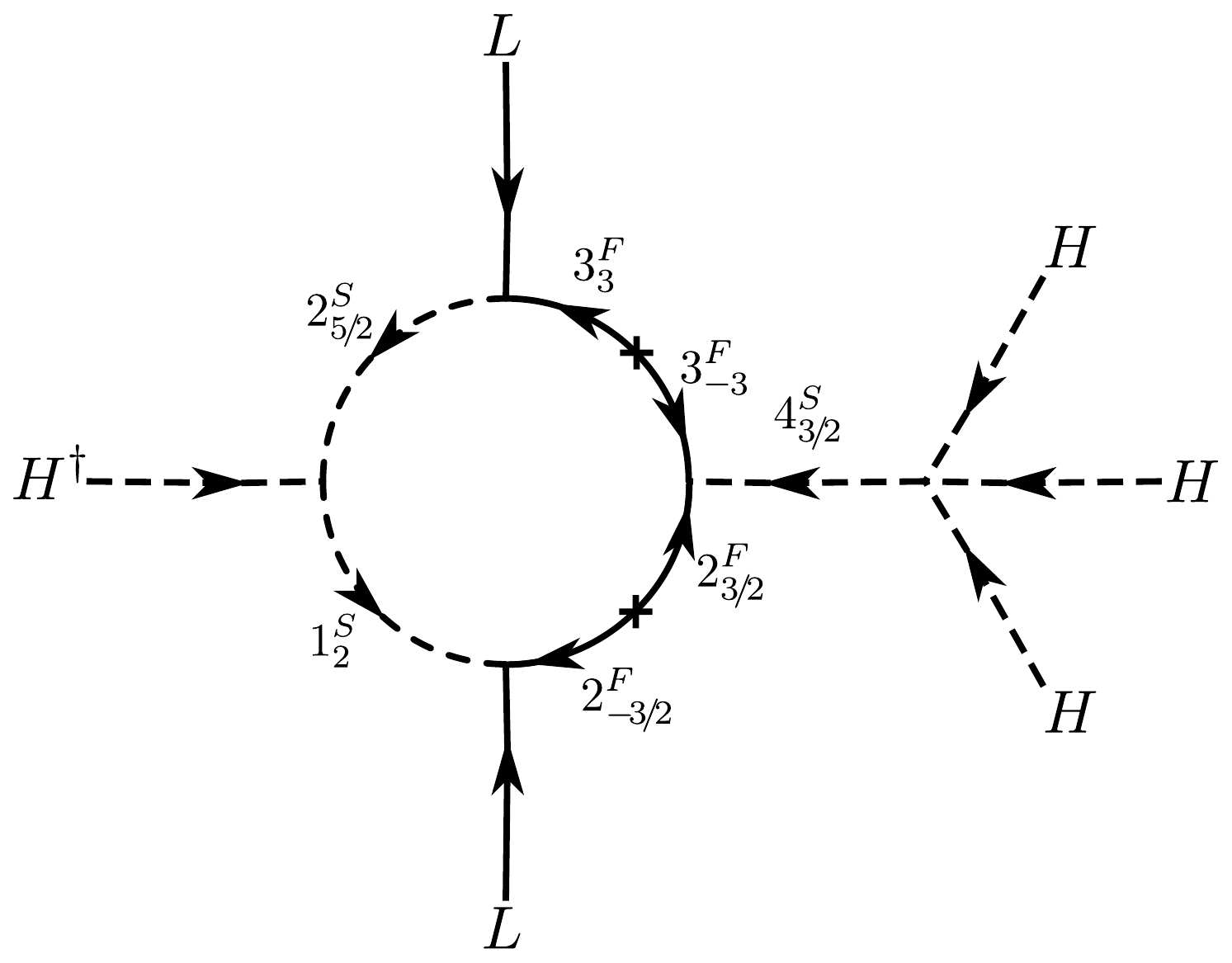}
  \caption{Two examples for genuine 1-loop $d=7$ diagrams. These
    examples have been chosen since they contain the smallest
    representations, for which genuine 1-loop $d=7$ diagrams can
    be constructed.}
  \label{fig:loopd7}
\end{figure}

At $d=7$ 1-loop one finds already 48 different topologies, from which,
however, only 8 can lead to genuine
models~\cite{Cepedello:2017eqf}. The analysis of
Ref.~\cite{Cepedello:2017eqf} shows that there is only one diagram in
which the largest internal representation can be as small as a
triplet, while there are a further 22 diagrams, with at least one
quadruplet. We will not repeat here all the diagrams for brevity and
instead show in Fig.~\ref{fig:loopd7} just two examples.

Figure~\ref{fig:loopd7} shows on the left the model with only
triplets, while the diagram on the right is an example of a model with
exactly one quadruplet. The exotic particles external to the loop in
these example models are ${\bf 3}_1^F$ (left diagram) and
${\bf 4}_{3/2}^S$ (right diagram). These two particles can not be
present in 1-loop $d=7$ models at the same time, otherwise one
generates the $d=7$ tree-level diagram of the BNT model. Again, as in
the case of $d=5$ 1-loop, one can build series of models allowing for
larger representations and/or hypercharges for the particles in the
loop. Also, from the fact that $H$ and $L$ have fixed quantum numbers,
one can derive a set of conditions on the possible combinations of
representations and hypercharges for the internal particles. The exact
conditions, however, depend on the type of diagram under
consideration. Note that most 1-loop $d=7$ diagrams need five new
particles (more, if one counts the vector partners of the fermions as
extra degrees of freedom), although for special values of the quantum
numbers four new fields are enough.

\subsection{Genuineness\label{subsect:gen}}

In this subsection we want to discuss our concept of ``genuineness''
for neutrino mass models in somewhat more detail. In short, we
consider a model genuine at dimension $d$, if all lower dimensional
contributions are automatically absent, without the need for
additional symmetries beyond those of the standard model group.

However, one aspect of higher-dimensional neutrino mass models needs
to be considered first. There is a single $\Delta L=2$ neutrino mass
operator of dimension $d$, which is always of the following
form:\footnote{Invariance under $SU(2)$ forces the operator dimension
  $d$ to be odd.  Note also that ${\cal O}^{d}$ in
  eq. \eqref{eq:defo7} contains $d-1\equiv 2n$ doublets of $SU(2)$, and in
  general the product of multiple doublets is expected to have many
  independent contractions. Indeed, it seems that the number of
  singlets in the product of $2n$ doublets is given by the Catalan
  numbers $C(n)\equiv(2n)!/\left[(n+1)!n!\right]$, hence for $d=13$
  (i.e. $n=6$) we might had expected 132 different contractions of the
  $SU(2)$ indices. Yet, we note that there is a single Higgs field, so
  all $(d-1)/2$ copies of $H$ must be contracted symmetrically; the
  same is true for the $(d-5)/2$ copies of $H^*$.  And it is not
  complicated to see that for a given $d$, there is always one---and
  only one---contraction with this property; all the others are
  therefore identically 0. See also Ref.~\cite{Liao:2010ku}.}
\begin{equation}\label{eq:defo7}
  {\cal O}^{d} \propto LLHH(H^{\dagger}H)^{(d-5)/2}\,,
\end{equation}
with the $SU(2)$ indices of each pair $LH$ outside the brackets
contracted with the anti-symmetric real tensor $\epsilon_{ab}$, and
each pair $H^\dagger H$ inside the brackets contracted with the
$\delta_{ab}$ tensor.

The very same operators will always lead to lower order loop
models:\footnote{Diagrammatically, this does not mean that one can
  close every pair of $H$,$H^*$ external lines. However, there will
  always be at least one such pairs of lines which can be closed.}
\begin{equation}\label{eq:nlp1}
  \frac{1}{\Lambda^{(d-4)}}LLHH(H^{\dagger}H)^{(d-5)/2} \rightarrow \frac{1}{16 \pi^2} 
  \frac{1}{\Lambda^{(d-6)}}LLHH (H^{\dagger}H)^{(d-7)/2}
\end{equation}
In the SM, where there is only one Higgs doublet, such loops
can not be forbidden by postulating some symmetry.
\footnote{For this reason, the authors of Ref.~\cite{Bonnet:2009ej}
  considered a two-Higgs doublet extension of the SM. Assigning
  different charges to the two doublets under a new $Z_n$ makes it
  possible to forbid loop contributions. Note, however, that these
  additional symmetries are spontaneously broken by the doublet vevs.}
One can straightforwardly estimate that such a loop contribution will
become more important than the tree-level one if $(\Lambda/v) \gsim
4\pi$.  This means $\Lambda \lsim 2$ TeV is required for the
$d$-dimensional tree-level contribution to dominate over the ($d-2$)
dimensional 1-loop one.  Since this is unavoidable in the SM, $d \ge
7$ tree-level model of neutrino mass {\em must have new particles
  below 2 TeV}, otherwise loop contributions will dominate the
neutrino mass matrix.  Note that this ``upper limit'' is more
stringent than the estimates for the typical scales $\Lambda$ shown in
Fig.~\ref{fig:scales}.

In loop calculations usually there appear both finite and infinite
loop integrals. However, in a renormalizable theory, infinite
contributions are canceled by counter-terms, implying that there are
lower order contributions to the same operator. Thus, all models with
diagrams requiring renormalization are not genuine in our sense. On
the other hand, diagrams associated to finite loop integrals only, can
lead to genuine models. One should distinguish two different
scenarios: Models in which lower order contributions are absent
automatically, and models which forbid lower order contributions with
the help of an extra symmetry. We consider only the former class of
models genuine.

Let us discuss the second scenario with one concrete and well-known
example: the scotogenic model~\cite{Ma:2006km}. Here, the right-handed
neutrino is assumed to be odd under a $Z_2$ and a new scalar doublet
(odd under the $Z_2$ as well) is added to the model. Thus, there is no
$d=5$ tree-level contribution from the SM Higgs and the 1-loop
contribution can dominate. The resulting 1-loop integral is finite and
thus, technically, no tree-level neutrino mass term is needed.  Let us
stress that while we do not consider such a construction to be
``genuine'' in our sense, such neutrino mass models are of course
perfectly valid and phenomenologically interesting models.

However, we also want to mention that such a construction relies on
the assumption that the new scalars in these models do not acquire a
vacuum expectation values. Of course, adding some discrete symmetry to
the model does not guarantee, by itself, the absence of a vev. Rather,
a non-zero vev for the exotic scalar(s) would break the discrete
symmetry spontaneously, leading to an unwanted tree-level neutrino
mass term and thus usually (but not always) vevs are to be avoided. This
can be achieved with an appropriate choice of parameter values in
the scalar potential.  (In the scotogenic model essentially it
corresponds to imposing the condition that the mass squared parameter
of the new scalar doublet is positive, $\mu_D^2 > 0$.)

\begin{figure}
  \centering
  \includegraphics[scale=0.45]{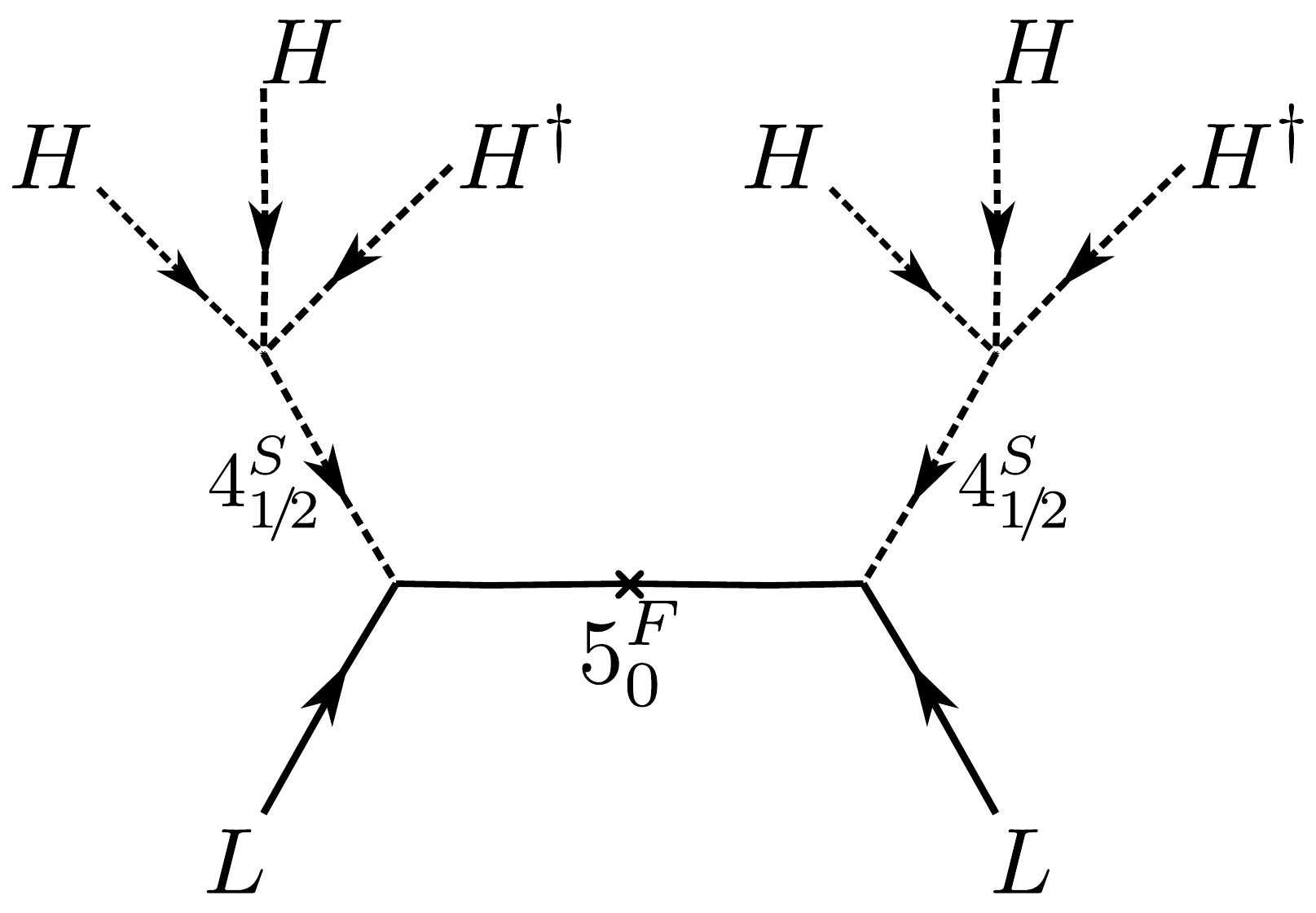}\hskip10mm
  \includegraphics[scale=0.45]{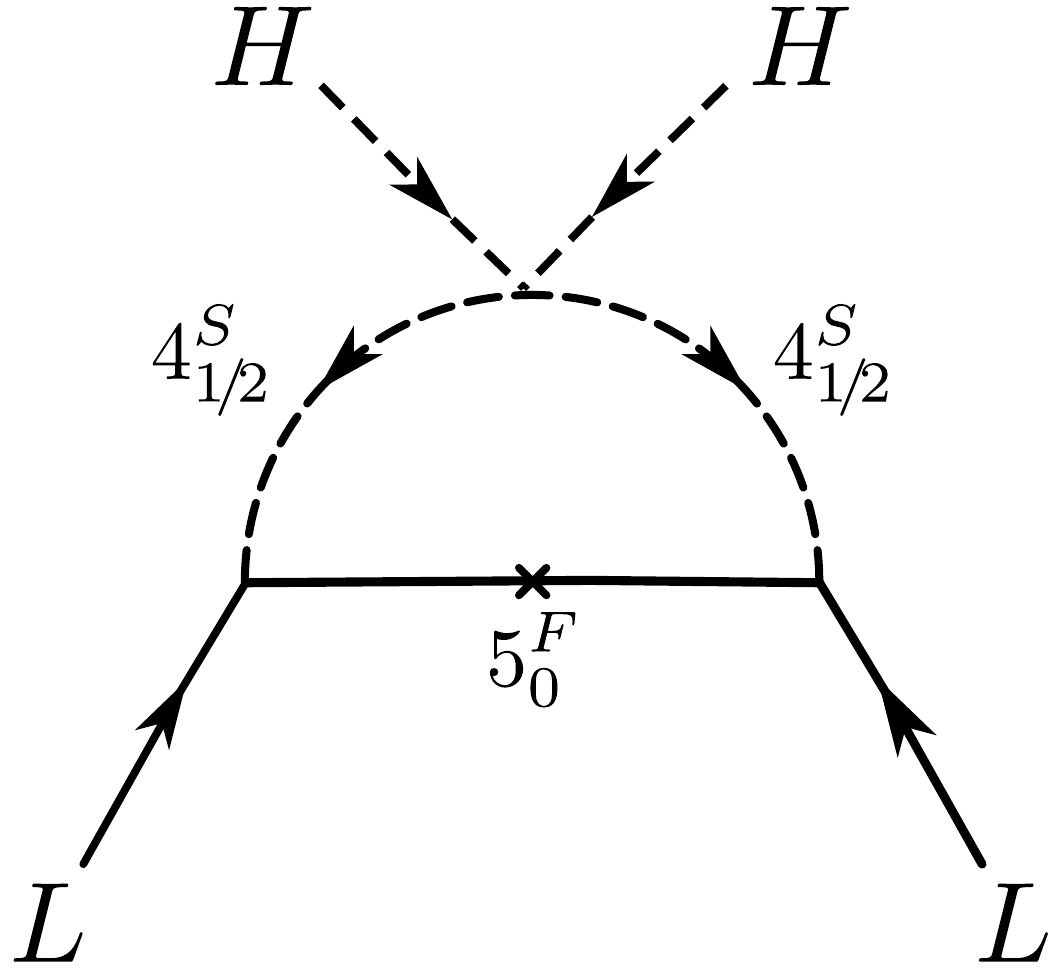}
  \caption{One example of a ``non-genuine'' $d=9$ diagram, to the left.
    Connecting the two quadruplet scalars, as shown on the right, leads
    to a 1-loop diagram at $d=5$.}
  \label{fig:d9ng}
\end{figure}

For concreteness, let us now consider a particular example model for a
$d=9$ tree-level diagram.  The model presented in Ref.~\cite{Kumericki:2012bh} 
contains two new fields: ${\bf 4}_{1/2}^S$ and
${\bf 5}_0^F$. This model is non-genuine in our definition.
\footnote{The same model was discussed also in Ref.~\cite{Nomura:2017abu}.}
The model generates a $d=9$ tree-level diagram, see
fig. (\ref{fig:d9ng}) on the left, via the four scalar vertex
$\lambda_4 ({\bf 4}_{1/2}^S)^{\dagger} H H H^{\dagger}$.\footnote{$\lambda_4$ 
  will lead to an induced vev for ${\bf 4}_{1/2}^S$
  even if the mass squared parameter $m_{S_{\bf 4}}^2$ is positive.}
Connecting the two quadruplet scalars via a quartic interaction
$\lambda_5 ({\bf 4}_{1/2}^S)^{\dagger} ({\bf 4}_{1/2}^S)^{\dagger} HH$
allows one to draw the 1-loop $d=5$ diagram on the right. The loop
integral is finite, just as in the scotogenic model. Assuming the
masses of ${\bf 4}_{1/2}^S$ and ${\bf 5}_0^F$ to be roughly of order
$\Lambda$ the ratio of the contributions of the two diagrams can be
estimated as
\begin{equation}
  \label{eq:rat}
  R({\rm tree/loop}) \propto \frac{\lambda_4^2}{\lambda_5}
  \Big(\frac{v_{SM}}{\Lambda}\Big)^4,
\end{equation}
i.e. the tree-level will be less important than the loop for scales
$\Lambda$ bigger than roughly $\Lambda \simeq 600
\sqrt{\lambda_4}/(\lambda_5)^{1/4}$ GeV.
Note that, since ${\bf 4}_{1/2}^S$ contains one doubly charged
component, the LHC searches on same-sign dileptons
\cite{ATLAS:2017iqw} should apply. Thus, one can estimate that the
current lower limits on the mass of ${\bf 4}_{++}^S$ should be in the
range of roughly [500,650] GeV, depending on the final state lepton
generation~\cite{Cepedello:2017lyo}.

Similar comments apply to the models presented in 
Refs.~\cite{Picek:2009is,Liao:2010cc,McDonald:2013kca,McDonald:2013hsa}.
Reference~\cite{Picek:2009is} introduces ${\bf 5}_1^F$, ${\bf 4}_{3/2}^S$ and
${\bf 4}_{1/2}^S$. The model has a a 1-loop diagram of type $T_3$, just
as in the example of Ref.~\cite{Kumericki:2012bh} discussed above.
Reference~\cite{Liao:2010cc} introduces the idea of a ``cascade
seesaw''. Essentially here the author discusses that models such as
\cite{Kumericki:2012bh} can be generalized to yield $d=9$, $d=13$ and
higher, by using larger and larger multiplets.
References~\cite{McDonald:2013kca,McDonald:2013hsa} discusses different seesaw
models at $d=7$ and $d=9$. However, this analysis considers only one
exotic fermion (and two new scalars) in each model. None of the
models in Refs.~\cite{McDonald:2013kca,McDonald:2013hsa} is genuine in our
sense.

\section{Classification and results\label{sect:class}}

The basic steps in the procedure are similar for $d=9$, $d=11$ and
$d=13$.  At each $d$ we first generate all possible topologies via a
computer code based on known algorithms --- for clear and self-contained explanations on how this can be done see~\cite{Read:1981,graphpaper2}. Here, we will only mention briefly the main idea. What we call a topology is what is known as a (undirected) graph to mathematicians, and it consists on some number of vertices connected among themselves by (undirected) edges. There is the obvious way of iteratively generating all graphs with $n$ vertices by adding one vertex to all graphs with one one less vertex, in all possible ways. However, this procedure generates many equivalent/isomorphic topologies. Checking whether two graphs are isomorphic is a well known problem, which can be time consuming. In order to avoid doing these checks as much as possible, one can instead use a more targeted recipe which generates all graphs with a given degree sequence (the degree of a vertex $i$ is the number of edges connected to it, and the ordered sequence of numbers $d_i$ constitutes the graph's degree sequence; for example the topologies in Fig. \ref{fig:d9ng} have degree sequences $443311111111$ and $4331111$). We used this latter approach \cite{Read:1981,graphpaper2}, even though a more naive one would probably also be feasible for the rather simple, tree-topologies under consideration.

Once these are obtained, we find all diagrams simply by labeling each
line as a fermion or a scalar in all possible ways, and ensuring that
one obtains fermion-fermion-scalar, scalar-scalar-scalar and scalar
four-point vertices only.  From these (large) lists of diagrams one
can construct all models by searching for every allowed combination of
$L$, $H$ and $H^{\dagger}$ in the outer legs of the diagrams.

From these lists we then eliminate every model, which is non-genuine
in our definition.  This is achieved in several steps, many of which can
be automatized (we have written a Mathematica code for this
purpose). The correctness of this code was cross-checked by visual
inspection for all $d=9$ and $d=11$ models. For $d=13$ we
cross-checked by hand that the models that were listed as genuine by
the code did indeed not lead to lower dimensional neutrino masses.
The basic idea behind this code is the following. Assume, for example,
that one starts with excluding models which do lead to tree-level
neutrino masses at $d=5$.  One could simply cross from the list of
models all those that contain the fields ${\bf 1}_0^F$, ${\bf 3}_0^F$
or ${\bf 3}_1^S$. To do so, however, it is not necessary to calculate
all the quantum numbers of the fields inside the diagrams. It is
sufficient to realize that, for example, if a combination of fields
such as $HL$ or $LL$ appear at the extreme end of a diagram, this is
equivalent to the existence of ${\bf 1}_0^F$/${\bf 3}_0^F$ or
${\bf 3}_1^S$ in the diagram. Similarly, all conditions for $d=7$ and
higher (and also for loops) can be reformulated as a search for
combination of fields that form certain groups after cuts into the diagrams.
We have found that programming these cuts is simpler than calculating
the quantum numbers for all fields and then eliminating unwanted
fields on a case-by-case basis.

We will not show all possible topologies and diagrams here
for brevity. However, the complete lists can be found at
\href{http://renatofonseca.net/high-dim-neutrino-masses.php}{renatofonseca.net/high-dim-neutrino-masses.php}.

\subsection{Dimension 9 ($d=9$)}

\begin{figure}
  \centering
  \includegraphics[width=.9\textwidth]{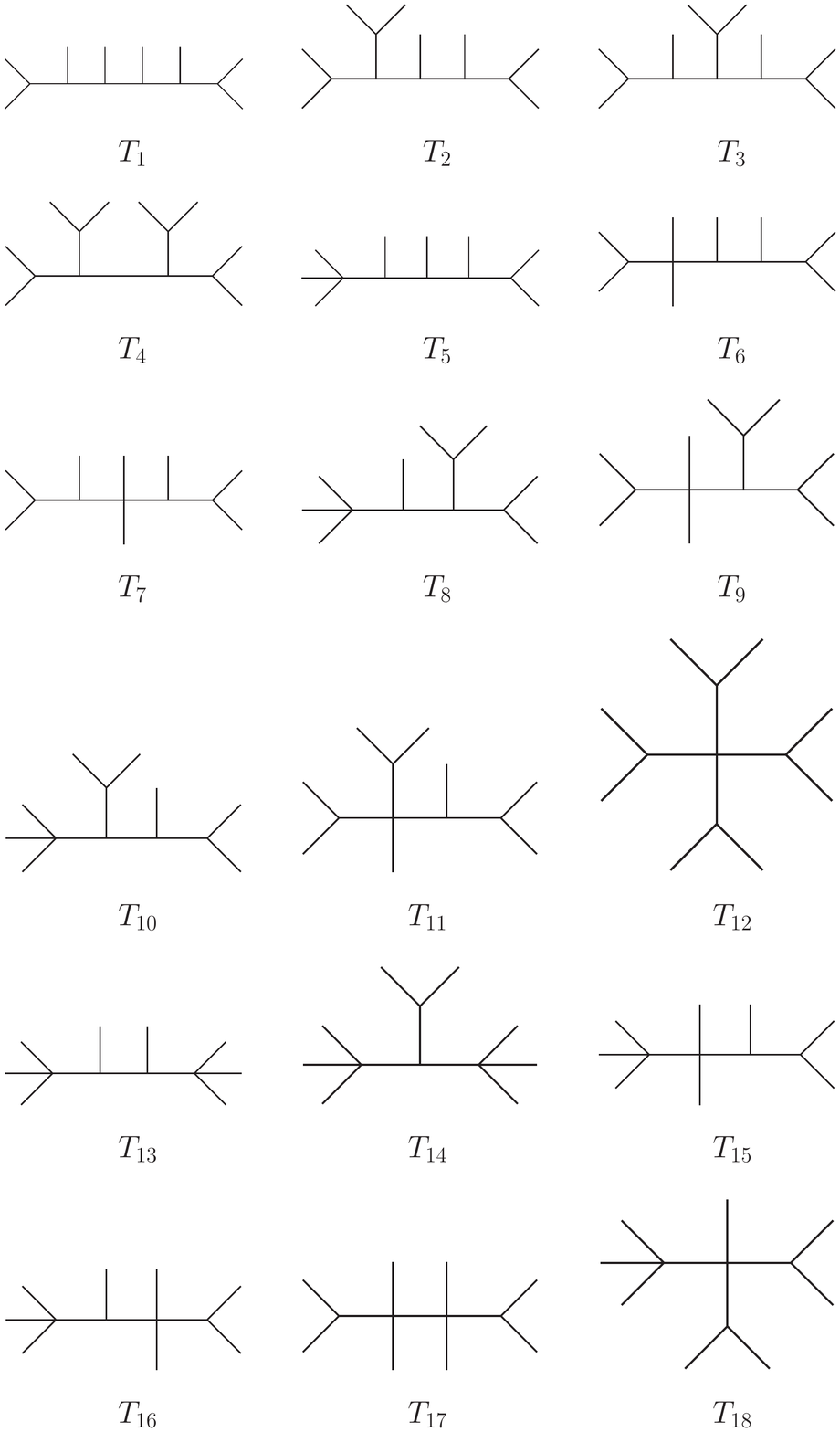}
  \caption{The 18 topologies at $d=9$ level that can give renormalizable
    diagrams.}
  \label{fig:d9topos}
\end{figure}

\begin{figure}
  \centering
  \includegraphics[scale=0.4]{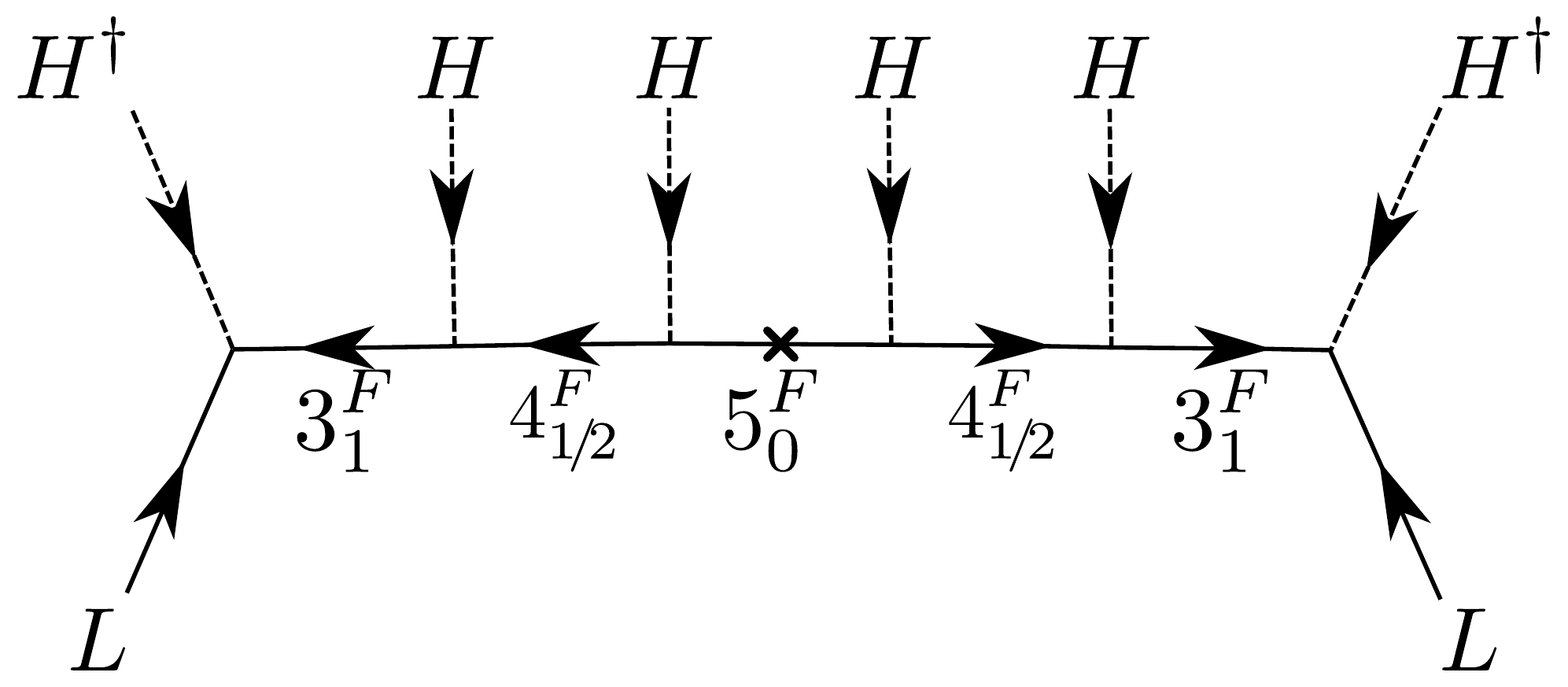}\hskip10mm
  \includegraphics[scale=0.4]{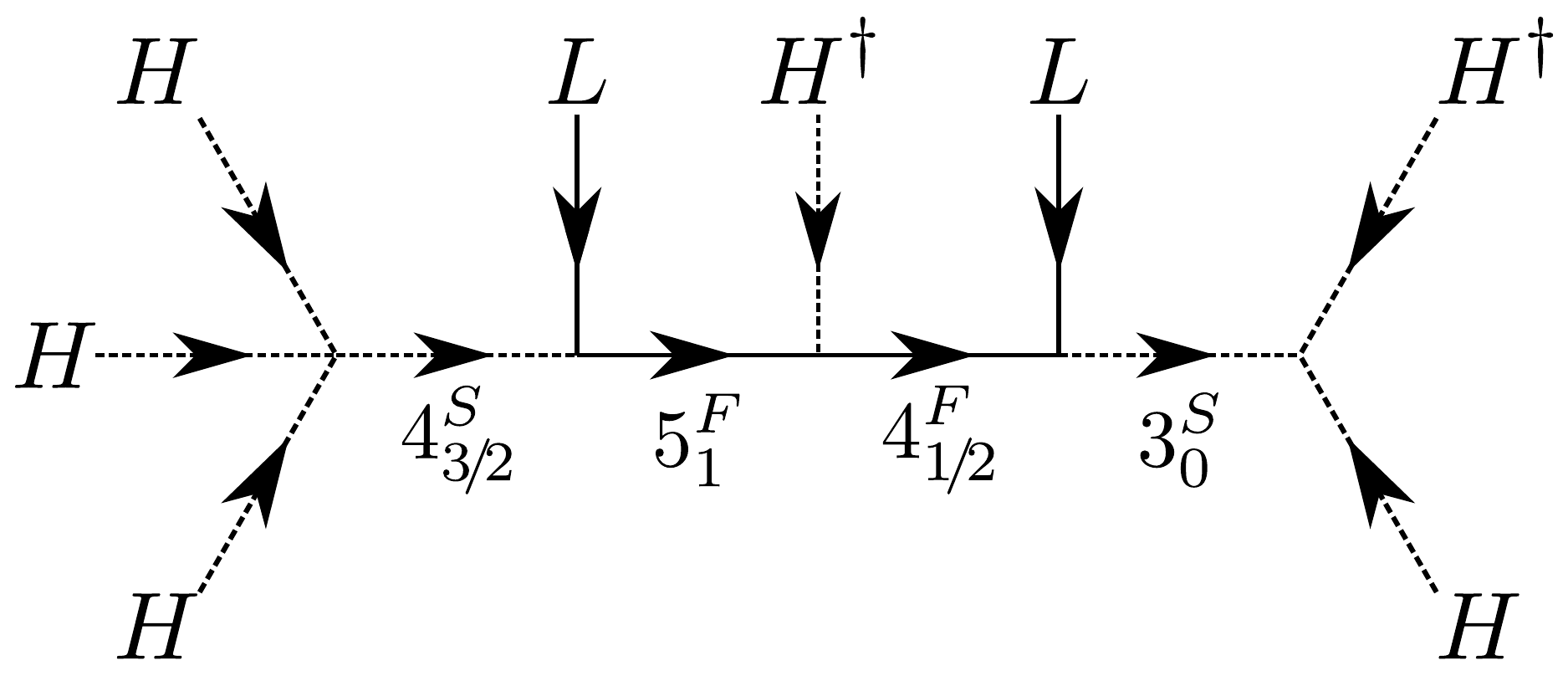}
  \caption{The two genuine diagrams at $d=9$.}
  \label{fig:d9diags}
\end{figure}

We start the discussion with $d=9$. Figure~\ref{fig:d9topos} shows all
18 topologies from which one can build valid neutrino mass diagrams
with renormalizable vertices only. There is one more topology (not
shown), with 8 external lines and no loops, but it requires three
4-point vertices, hence it will lead only to non-renormalizable
models.  The 18 topologies which we do show generate a total of 66
diagrams. However, all except four topologies lead only to diagrams
that necessarily have a tree-level neutrino mass at either $d=5$ or
$d=7$. Diagrams from two more topologies will always also generate
1-loop $d=5$ diagrams hence, in the end, only topologies $T_1$ and
$T_5$ yield diagrams that are genuine in our sense. But not all
diagrams obtained from $T_1$ and $T_5$ are genuine either; the only ones which
are genuine can be seen in Fig.~\ref{fig:d9diags}.

Consider first the diagram on the left hand side of
Fig.~\ref{fig:d9diags}.  It contains only three new fermions, ${\bf
  3}_1^F$, ${\bf 4}_{1/2}^F$ and ${\bf 5}_0^F$, together with their
vector partners, and no exotic scalar. This is the minimal genuine
model at $d=9$. Its Lagrangian is given by:
\begin{equation}
  \label{eq:lagd9m1}
        {\cal L} = {\cal L}_{SM}+ {\cal L}_{Yuk} + {\cal L}_{mass},
\end{equation}
where
\begin{equation}
  \begin{aligned}
    \label{eq:yuk9m1}
          {\cal L}_{Yuk} &= Y_{\nu} \, L \cdot {\bf 3}_{1}^F H^{\dagger} 
          +  {\overline Y}_{\bf 34} \, {\bf 3}_{-1}^F \cdot  {\bf 4}_{1/2}^F H
          +  Y_{\bf 34} \, {\bf 3}_{1}^F\cdot  {\bf 4}_{-1/2}^F  H^{\dagger}
          \\ 
          & \quad +  {\overline Y}_{\bf 45} \, {\bf 4}_{1/2}^F \cdot  {\bf 5}_{0}^F  H^{\dagger}
          +   Y_{\bf 45}  \, {\bf 4}_{-1/2}^F \cdot {\bf 5}_{0}^F   H
  \end{aligned}
\end{equation}
and
\begin{equation}
  \label{eq:mass9m1}
        {\cal L}_{mass} = M_{\bf 3} {\bf 3}_{1}^F{\bf 3}_{-1}^F
        + M_{\bf 4} {\bf 4}_{1/2}^F{\bf 4}_{-1/2}^F
        + M_{\bf 5} {\bf 5}_{0}^F{\bf 5}_{0}^F .
\end{equation}
The light neutrino mass can be estimated in seesaw approximation
(for one generation) as:
\begin{equation}
  \label{eq:numass1}
  m_{\nu} \simeq Y_{\nu}^2 {\overline Y}_{\bf 34}^2 Y_{\bf 45}^2
  \frac{v_{SM}^6}{M_{\bf 3}^2 M_{\bf 4}^2M_{\bf 5}}
\end{equation}
For masses of the order of ${\cal O}(1-2)$ TeV, Yukawas of the order
of $(0.03-0.04)$ will reproduce the scale of the atmospheric neutrinos,
$m_{\nu} \simeq \sqrt{\Delta(m^2_{Atm})} \simeq 0.05$ eV. For a more
detailed fit of neutrino masses and angles, see the appendix.

The diagram on the right hand side of Fig.~\ref{fig:d9diags} contains two
exotic scalars and two exotic fermions. We give only the part of
the Lagrangian relevant for the calculation of the neutrino mass,
\begin{equation}
  \begin{aligned}
    \label{eq:d9Lag2}
          {\cal L} & \propto  \lambda_4 H H H ({\bf 4}_{3/2}^S)^{\dagger}
          + \mu_3 {\bf 3}_0^S H H^{\dagger} + m_4^2 |{\bf 4}_{3/2}^S|^2
          +  m_3^2 |{\bf 3}_0^S|^2
          \\ 
          & \quad + Y_{5L} L \cdot {\bf 5}_{-1}^F {\bf 4}_{3/2}^S
          + Y_{45} {\bf 4}_{-1/2}^F \cdot {\bf 5}_1^F H^{\dagger}
          + Y_{4L} L \cdot {\bf 4}_{1/2}^F ({\bf 3}_0^S)^{\dagger}
          \\ 
          & \quad + M_{\bf 5_1} {\bf 5}_{-1}^F{\bf 5}_{1}^F
          + M_{\bf 4} {\bf 4}_{-1/2}^F{\bf 4}_{1/2}^F.
  \end{aligned}
\end{equation}
Again in seesaw approximation and for one generation we can roughly
estimate the size of the neutrino mass generated by this model
as,
\begin{equation}
  \label{eq:numass2-9}
  m_{\nu} \simeq Y_{5L} Y_{4L} Y_{45}\lambda_4
  \frac{\mu_ 3v_{SM}^6}{M_{\bf 5_1} M_{\bf 4}m_ 4^2m_3^2}.
\end{equation}
With all mass parameters equal to 1 TeV, $\mu_3=M_{\bf 5_1}= M_{\bf4}
=m_4=m_3=1$ TeV, and for $\lambda_4=Y_{5L}= Y_{4L} =Y_{45}={\cal O}(10^{-2})$
this gives roughly $0.3$ eV. A more detailed description on how all
neutrino data can be fitted in this model is deferred to the appendix.

\subsection{Dimension 11 ($d=11$)}

\begin{figure}
  \centering
  \includegraphics[scale=0.4]{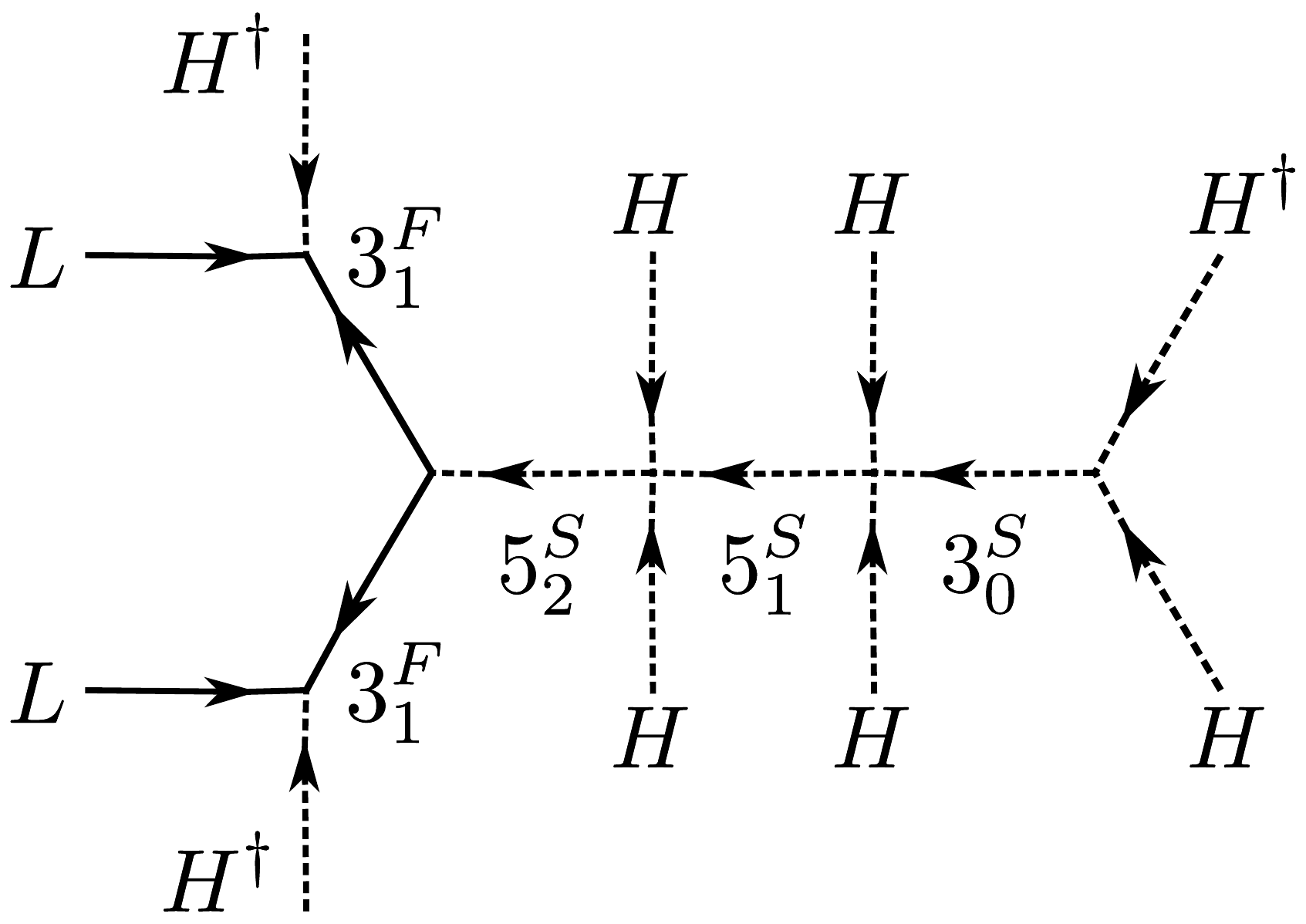}\hskip10mm
  \includegraphics[scale=0.4]{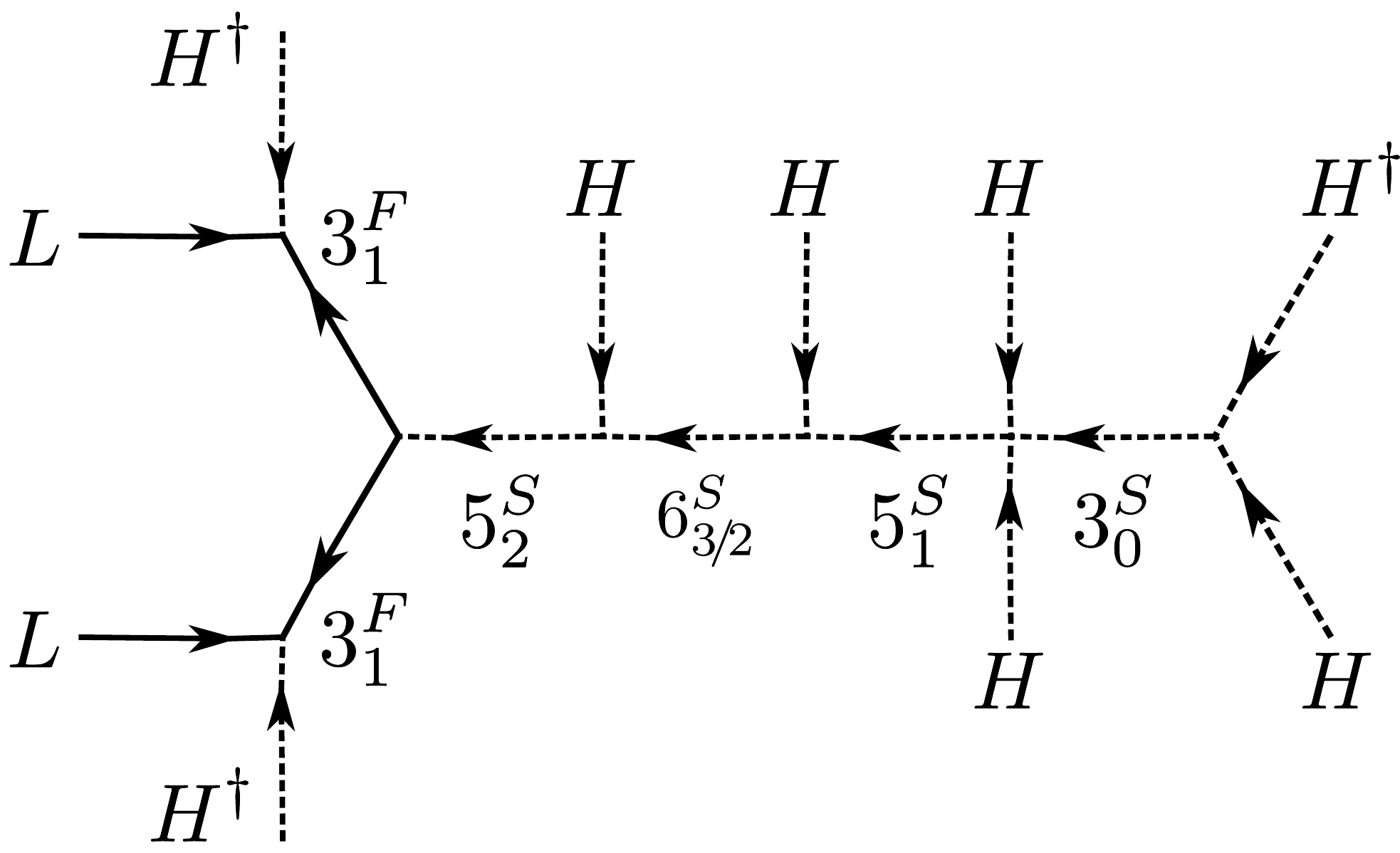} 
  \caption{The two genuine diagrams at $d=11$. Note that the fields on
    the right-hand side diagram always produce the diagram on the
    left-hand side.}
  \label{fig:d11diags}
\end{figure}

At $d=11$ we find 92 topologies, which generate a total of 504
diagrams.  It is not very instructive to discuss in detail all the
topologies and diagrams, as the methodology for eliminating
non-genuine models is the same as for the $d=9$ case. The only
two genuine diagrams are shown in Fig.~\ref{fig:d11diags}, and they are based
on very similar models. The diagram on the right contains five new particles,
four of which are also present in the diagram on the left. Thus, the
model for the right diagram always produces also the diagram on the
left. Unsurprisingly, at $d=11$ genuine diagrams require at least four
different beyond-SM particles and large representations: At least two
different quintuplets are needed and the model shown in the right diagram
of Fig.~\ref{fig:d11diags} requires in addition a sextuplet.

Again, we write down only the part of the Lagrangian relevant for
estimating the neutrino mass,
\begin{equation}
  \begin{aligned}
    \label{eq:d11Lag}
          {\cal L} & \propto \lambda_{55}{\bf 5}_{-2}^S{\bf 5}_1^S H H
          + \lambda_{53}{\bf 5}_{-1}^S{\bf 3}_0^S H H
          + \mu_{61}{\bf 5}_{1}^S{\bf 6}_{-3/2}^S H
          + \mu_{62}{\bf 5}_{-2}^S{\bf 6}_{3/2}^S H
          + \mu_3 {\bf 3}_0^S H H^{\dagger}
          + m_3^2 |{\bf 3}_0^S|^2 
          \\ 
          & \quad  + m_{5_1}^2 |{\bf 5}_{1}^S|^2
          + m_{5_2}^2 |{\bf 5}_{2}^S|^2
          + m_{6}^2 |{\bf 6}_{3/2}^S|^2
          + Y_{\nu} L \cdot {\bf 3}_{1}^F  H^{\dagger}
          + Y_{3} {\bf 3}_{-1}^F \cdot {\bf 3}_{-1}^F {\bf 5}_2^S
          + M_{\bf 3} {\bf 3}_{-1}^F{\bf 3}_{1}^F.
  \end{aligned}
\end{equation}
A simple estimate for the neutrino mass from the left diagram
in Fig.~\ref{fig:d11diags} gives:
\begin{equation}
  \label{eq:numass1-11}
  m_{\nu} \simeq Y_{\nu}^2 Y_{3} \lambda_{55}\lambda_{53}
  \frac{\mu_ 3v_{SM}^8}{M_{\bf 3}^2 m_3^2  m_{5_2}^2  m_{5_2}^2 }
\end{equation}
For a new physics scale of $\Lambda=1$ TeV and all dimensionless couplings
order $0.05$ one finds again a neutrino mass of order $0.3$ eV. As is
also the case for the second of our $d=9$ models, the dimensionful
scalar coupling $\mu_3$ can be a source of additional neutrino mass
suppression. For $\mu \simeq 100$ keV and all dimensionless couplings
${\cal O}(1)$ ($\Lambda=1$ TeV) one finds $m_{\nu} \sim {\cal O}(0.1)$ eV.

A straightforward calculation shows that the diagram on the right
gives a neutrino mass of roughly the same numerical value, if
$(\mu_{61} \mu_{62})/m_6^2 \simeq \lambda_{55}$. Thus, the diagram on
the left is the dominant one, if either $\mu_{61}$ or $\mu_{62}$ (or
both) are very small relative to the new physics scale $\Lambda$. On
the other hand, for $\lambda_{55} \ll 1$ the opposite situation
is found.

\subsection{Dimension 13 ($d=13$)}

At $d=13$ there are 576 topologies and 4199 diagrams. One has to
delete not only all models that lead to a tree level $d=5$, $d=7$,
$d=9$ and $d=11$ diagram, as well as a 1-loop $d=5$ or $d=7$ diagram,
but also all models with a 1-loop $d=9$ diagram.  The last cut
drastically reduces again the list of genuine models: without it
nearly 50 different diagrams remain, while after this cut only 2
genuine tree-level diagrams at $d=13$ remain.

\begin{figure}
  \centering
  \includegraphics[width=0.75\textwidth]{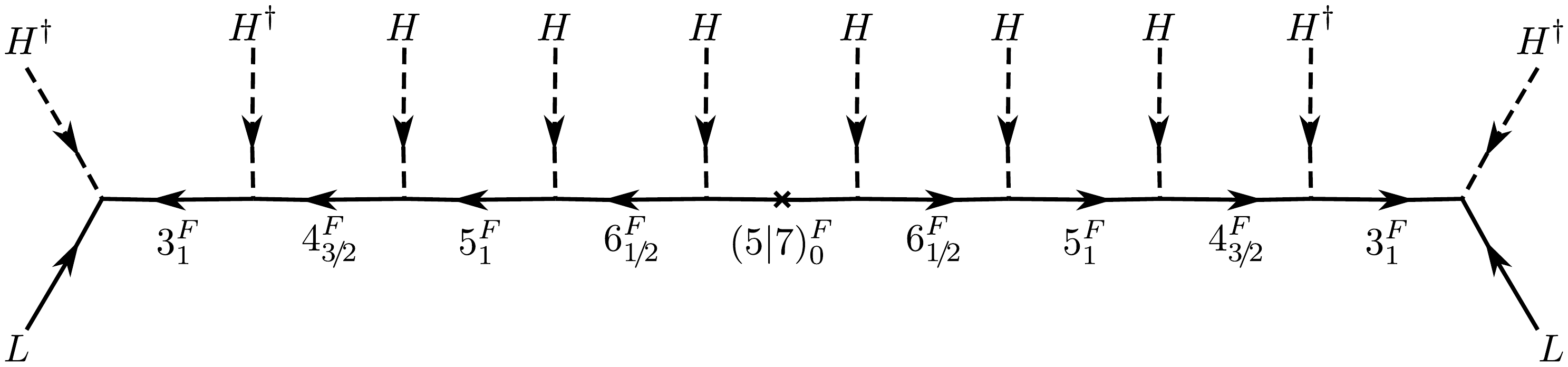}
  \\
  \includegraphics[width=0.75\textwidth]{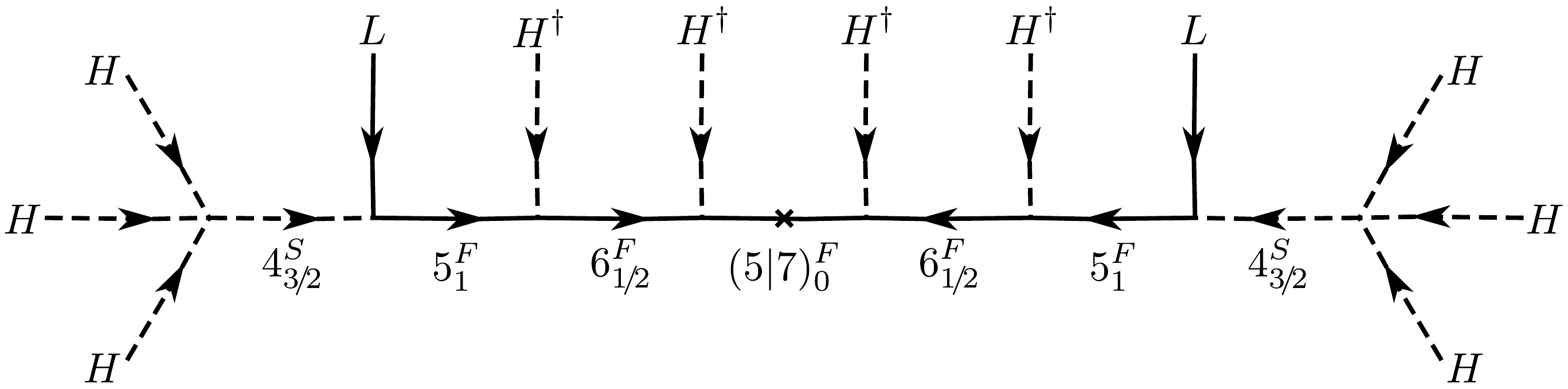} 
  \caption{The two genuine diagrams at $d=13$ can be realized in a
    total of 6 models. It is possible to have either a ${\bf 5}_{0}^F$
    or a ${\bf 7}_{0}^F$ in the middle.  Furthermore, two extra models
    are obtained from the top diagram by rearranging appropriately the
    external $H$'s and $H^{\dagger}$'s and replacing either one or both of the
    internal ${\bf 4}_{3/2}^F$ correspondingly by a
    ${\bf 4}_{1/2}^F$. Note also that these two extra models (not
    represented above) with the field ${\bf 4}_{1/2}^F$ need ${\bf
      7}_{0}^F$ in the middle of the diagram in order to be genuine.}
  \label{fig:d13diags}
\end{figure}

Figure~\ref{fig:d13diags} shows the two remaining genuine diagrams.
Unsurprisingly, more fields and larger representations are needed in
these diagrams. The largest representation is now a $SU(2)_L$
septet. There is a total of six model variations that one can find
for these two diagrams. In addition to the four models shown, one
can construct two more model variations for the first diagram (top
row): Replace either one or both of the ${\bf 4}_{3/2}^F$ by a
${\bf 4}_{1/2}^F$ (rearranging $H$ and $H^{\dagger}$
correspondingly). Note that the models with ${\bf 4}_{1/2}^F$
are only genuine with a ${\bf 7}_{0}^F$.
 
Let us discuss first briefly the models corresponding to the diagram
in the top row.  These models contain only new fermions, but no 
exotic scalars, and are very similar to each other. The models shown
contain five new fermions. As mentioned above, there are two more
variations containing a ${\bf 4}_{1/2}^F$. Comparing these
fermion-only models with the simplest $d=9$ model, one sees that
higher dimensional fermion-only diagrams ($d=17,21$ etc.) could be
straightforwardly found, following the same construction principles.

We will not write down the complete Lagrangian for these $d=13$ models
for brevity. The neutrino mass is estimated for these models to be of
order $m_{\nu} \simeq Y^{10}\frac{v^{10}}{\Lambda^9}$, where $Y$
stands symbolically for the Yukawa couplings in the diagrams and we
assumed for simplicity that all masses are of order $\Lambda$. Yukawa
couplings now have to be of order ${\cal O}(0.3)$ (with $\Lambda=1$
TeV) for a neutrino mass $m_{\nu} \simeq (0.1-0.2)$ eV.

The remaining $d=13$ models in the bottom row of Fig.~\ref{fig:d13diags}
need four exotic fields, one of them needs to be an 
exotic scalar. Again, a fermionic septet is the largest $SU(2)_L$
representation. Since in these models, some of the Yukawa couplings
from the fermion-only models are replaced by four-point scalar
couplings, slightly smaller couplings, say ${\cal O}(0.2)$, are
needed here to achieve $m_{\nu} \simeq (0.1-0.2)$ eV.  We close this
subsection by stating again that all $d=13$ models can easily fit all
measured neutrino mass squared differences and angles.

\subsection{A short discussion of phenomenological aspects}
\label{subsect:Constr}

In this paper we are mostly concerned in classifying neutrino mass
models. For completeness, however, in this section we add a brief
discussion of the phenomenology of these models. For brevity, we
will focus on only two aspects. First we discuss the running of the
gauge couplings. Then, we turn to collider physics.

\subsubsection{Running of gauge couplings}
\label{subsubsect:RGE}

A common feature of all genuine tree-level neutrino models identified in  the previous
sections is that large $SU(2)_L$ multiplets are required in order to
avoid lower dimensional contributions to the neutrino mass matrix.
Adding new $SU(2)_L$ multiplets to the standard model changes the
running of the gauge couplings. Since our tree-level models do not
add any beyond the standard model (BSM) coloured fields, the
value of the  $SU(3)_c$ coupling constant is not affected. 

\begin{figure}
  \centering
  \includegraphics[width=0.75\textwidth]{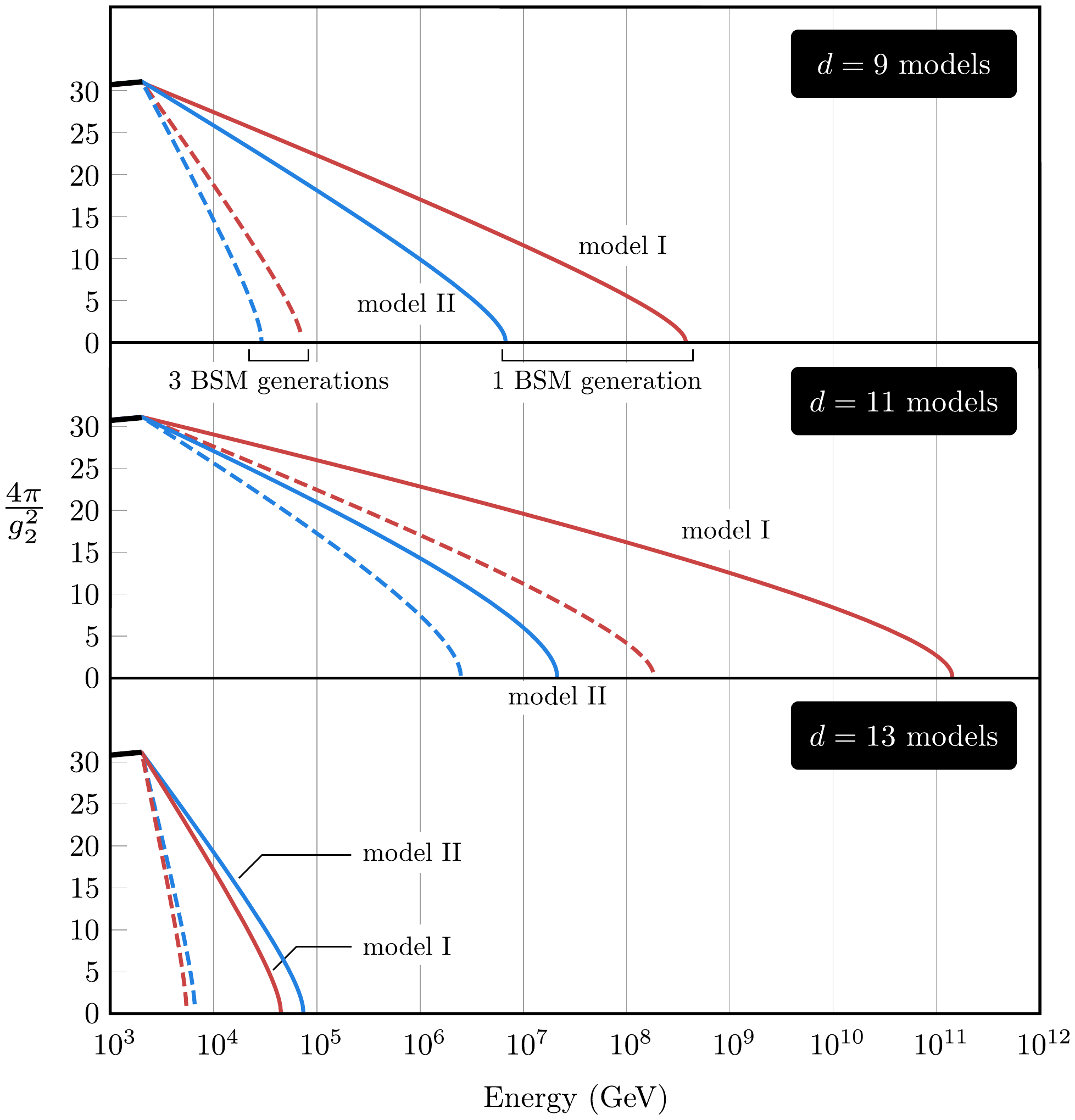}
  \caption{Running of the (inverse of) the gauge coupling $g_2$ for six of the eight
    different genuine neutrino mass models we have found: two at $d=9$, plus two at $d=11$ and six at $d=13$ (in this last case, we do not show the running for the two models with a fermion septet). For a summary of the fields in each model shown here, see appendix \ref{subsect:RGE}. Continuous lines are for a single copy of each exotic fermion representation, while the dashed lines show the running of $g_2$ for three families. In all cases
  Landau poles appear below the Planck scale, assuming a new physics scale of 2 TeV.}
  \label{fig:rge}
\end{figure}

Fig. (\ref{fig:rge}) shows the running of the inverse of $g_2^2$ as a
function of energy for six different models. These are the two+two genuine
models with $d=9$ and $d=11$, plus two more with $d=13$. Note that we found six $d=13$ genuine neutrino mass models: the two considered in
fig. (\ref{fig:rge}) contain the multiplets $5_F^0$ (the variants with
$7_F^0$ exhibit an even stronger running of $g_2$). In appendix
(\ref{subsect:RGE}) we give the $\beta$ coefficients at 1-loop and
2-loop for these models. The numerical results shown in fig.
(\ref{fig:rge}) assume that the new states have all masses of roughly
$m \sim 2$ TeV. For each model we show two different cases: (i) the
running when only one copy of the exotic fermions is present and (ii)
with three copies of exotic fermions. In all cases we use only one
copy for the exotic scalars. This choice is motivated by the fact that
standard model fermions come in three generations, but the neutrino
mass fit in several of the models could be done with only one copy
of the exotic fermions. Running is stronger, of course, with more fermions.
Thus, one can understand our choice as representing the extreme
cases, that could be realized.

As the figure shows, for all models Landau poles appear in $g_2$ below
the grand unification scale (GUT). This affirms again our supposition
that all these high dimensional neutrino mass models really have to be
considered as low-energy constructions. The energy at which the Landau
poles appear varies strongly from model to model. For $d=13$ and three
generations, the blow-up of the gauge coupling would occur as low
as 10 TeV. Finally, we mention that we have checked that in all cases
the running of $g_Y$ is less strong, than the one we show for $g_2$.
For this reason, only $g_2$ is shown in fig. (\ref{fig:rge}).

\subsubsection{Collider physics}
\label{subsubsect:LHC}

A detailed study of all possible production and decay channels for
each of the models, presented above, is beyond the scope of this
paper. Instead, here we will concentrate on the most promising signals
for the exotic fermions. This choice is motivated by the fact that
{\em all} models discussed do introduce some exotic fermions, while
exotic scalars appear only in some of the models.  We will add,
however, some comments on exotic scalars at the end of this section.

We have implemented our first $d=9$ model in SARAH
\cite{Staub:2012pb,Staub:2013tta}. Using Toolbox \cite{Staub:2011dp},
the implementation can be used to generate SPheno code
\cite{Porod:2003um,Porod:2011nf}, for the numerical evaluation of mass
spectra and observables.  Production cross sections for the different
fermions are then calculated using MadGraph \cite{Alwall:2014hca}.

\begin{figure}
  \centering
  \includegraphics[width=0.3\textwidth]{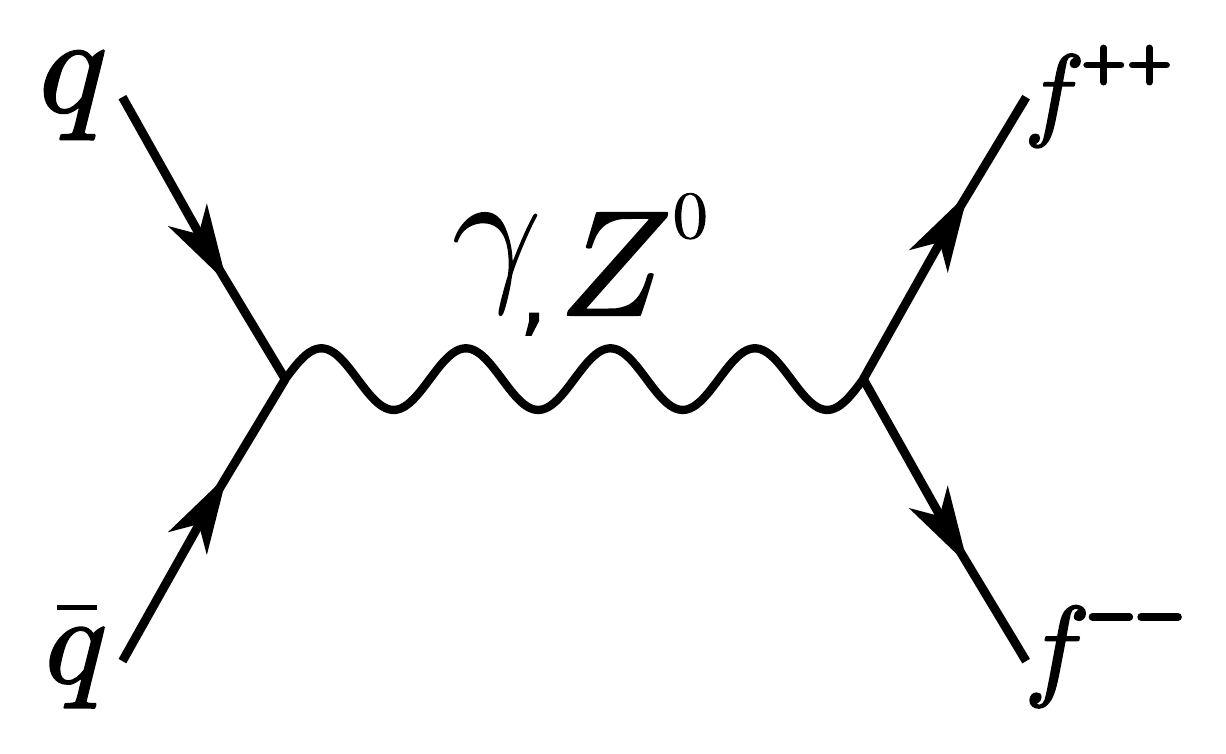}\hskip10mm
  \includegraphics[width=0.3\textwidth]{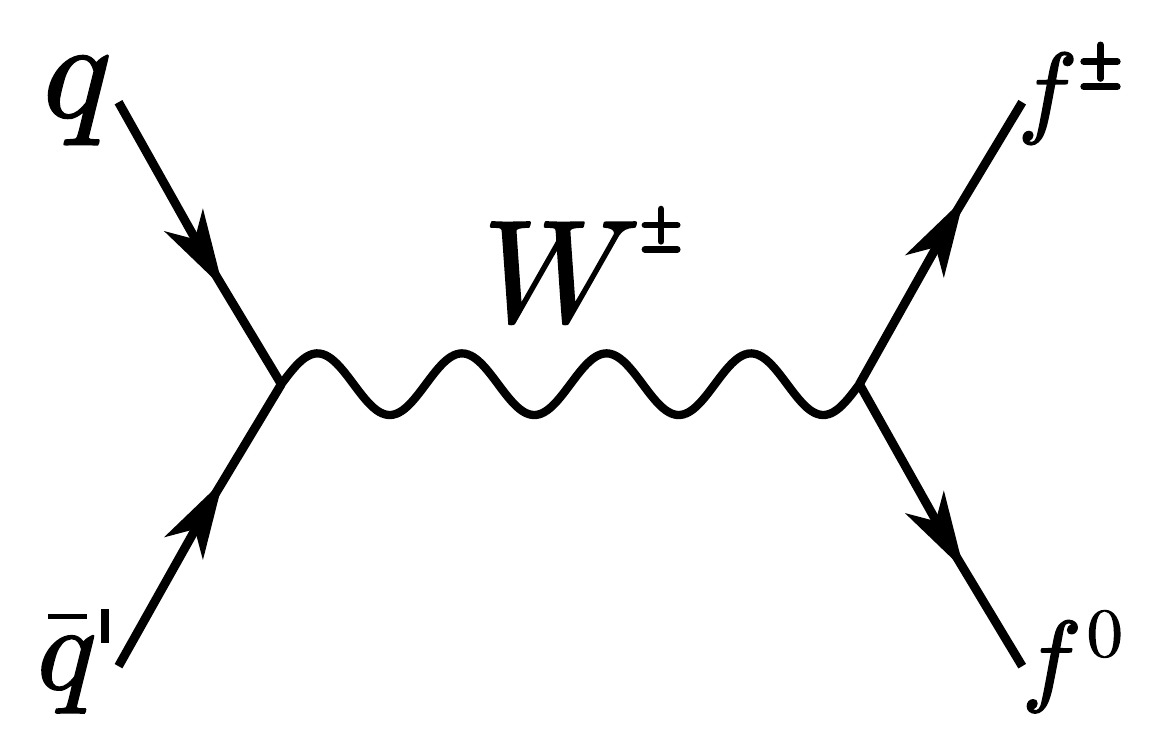}
  \caption{Example Feynman diagrams for the production of exotic
    fermions appearing in our high-dimensional neutrino mass models.
    Here, $f$ stands symbolically for a fermion from the multiplets
    ${\bf 3}^F_{1}$, ${\bf 4}^F_{1/2}$, ${\bf 4}^F_{3/2}$, ${\bf 5}^F_0$, ${\bf 5}^F_{1}$, ${\bf 6}^F_{1/2}$, ${\bf 6}^F_{3/2}$ or ${\bf 7}^F_0$.}
  \label{fig:prod}
\end{figure}

\begin{figure}
  \centering
  \includegraphics[width=0.45\textwidth]{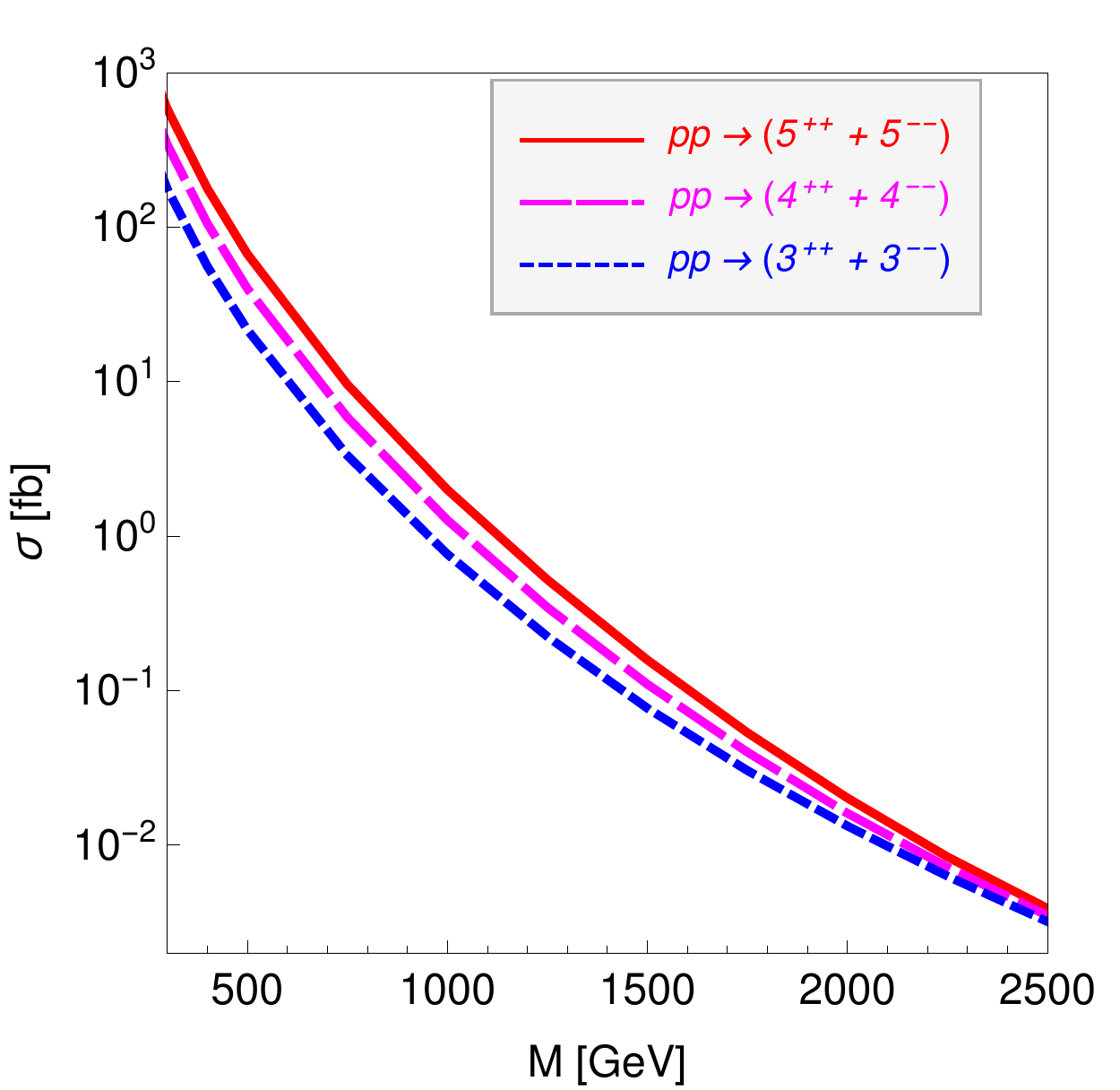}\hskip8mm
  \includegraphics[width=0.45\textwidth]{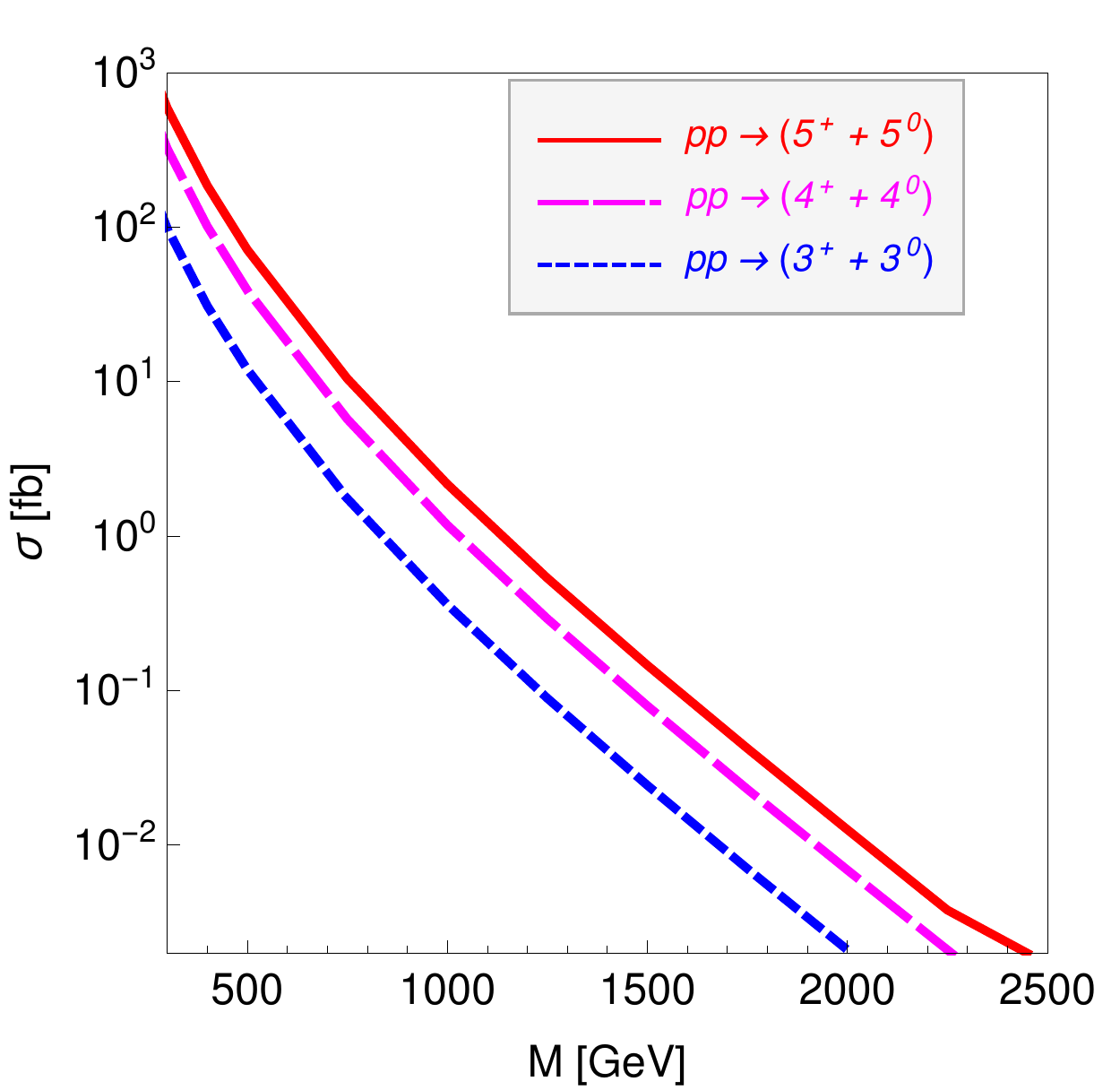}
  \caption{Cross sections for pair production of doubly charged
    fermions at the LHC for $\sqrt{s}=13$ TeV, to the left. To the
    right, associated production of fermions, $f^+f^0$ for $f$ from
    ${\bf 3}^F_{1}$, ${\bf 4}^F_{1/2}$ or ${\bf 5}^F_0$.}
  \label{fig:xsect}
\end{figure}

As we have argued in section (\ref{subsect:gen}), the new particles of
our high-dimensional neutrino mass models should not be heavier than
very roughly $ m \sim 2$ TeV, otherwise the tree-level diagram(s) will
not give the dominant contribution to the neutrino mass matrix.
A natural question to ask then is, whether LHC searches will be able
to cover this mass range completely, once sufficient luminosity has
been accumulated.

Fig. (\ref{fig:prod}) shows some example Feynman diagrams for
production of exotic fermions. Here, $f$ stands symbolically for any
fermion from the multiplets ${\bf 3}^F_{1}$, ${\bf 4}^F_{1/2}$, ${\bf 4}^F_{3/2}$, ${\bf 5}^F_0$, ${\bf 5}^F_{1}$, ${\bf 6}^F_{1/2}$, ${\bf 6}^F_{3/2}$ or ${\bf 7}^F_0$.  Numerical example cross sections for pair production of
$f^{++}f^{--}$ and associated production of $f^{+}f^{0}$ are shown in
fig. (\ref{fig:xsect}) as a function of the fermion mass. These cross
sections are for the LHC with a $\sqrt{s}=13$ TeV. The cross sections
are calculated for ${\bf 3}^F_{1}$, ${\bf 4}^F_{1/2}$ and ${\bf 5}^F_0$, assuming there is
no mixing between the multiplets.  Cross sections for fermions from
the multiplets $4_{3/2}^F$ and $5_1^F$ are slightly larger than the
ones shown in the figure. Note that the smallest cross sections are
found in all cases for $3_1^F$.

The plots show that the cross sections for pair production and
associated productions are similar for smaller values of the masses,
while the pair production cross sections are much larger than
associated production for the largest values of masses shown.  For
this reason, one expects in general more stringent constraints on
these fermions will come from searches for pair produced fermions.  As
fig. (\ref{fig:xsect}) shows, pair production cross sections are nearly
independent of the $SU(2)_L$ quantum numbers for the largest masses.  In
all cases, the pair production cross section is larger than $10^{-2}$
fb for masses up to $m =2$ TeV. Recall that this corresponds to
roughly 30 events for the high-luminosity LHC (HL-LHC), albeit before
cuts.

\begin{figure}
  \centering
  \includegraphics[width=0.3\textwidth]{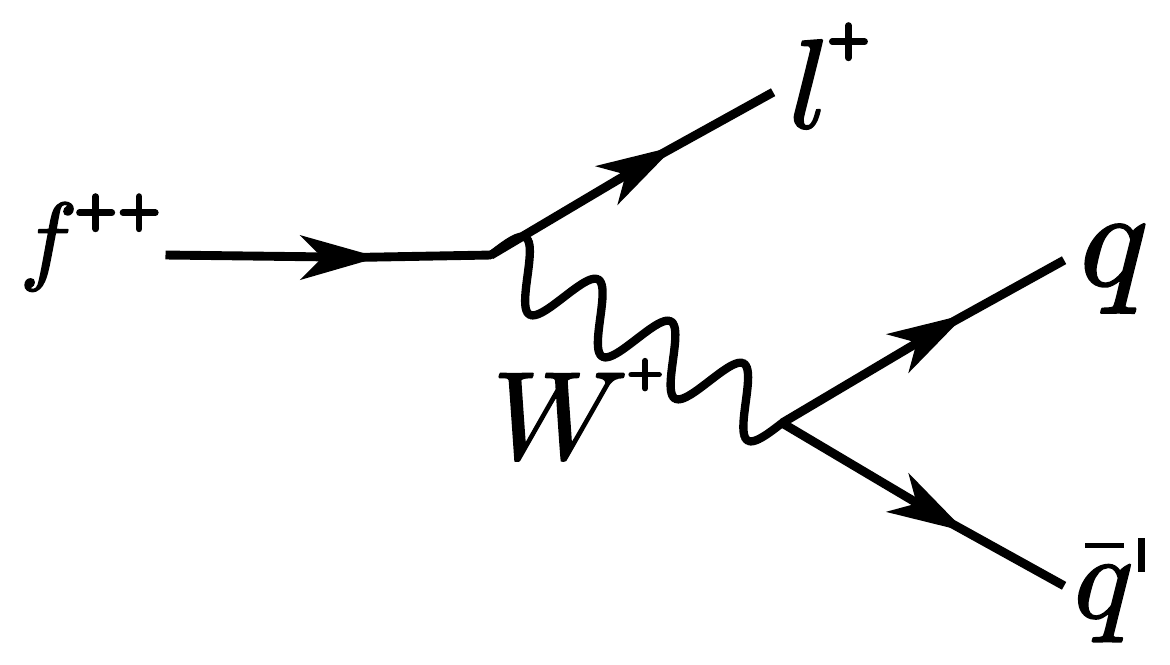}\hskip4mm
  \includegraphics[width=0.3\textwidth]{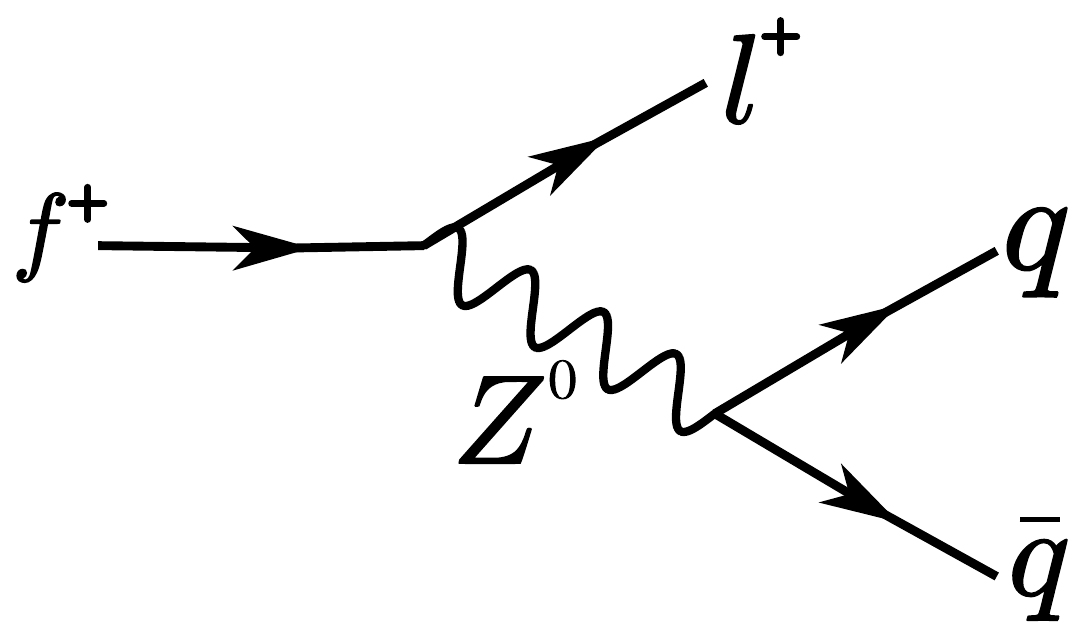}\hskip4mm
  \includegraphics[width=0.3\textwidth]{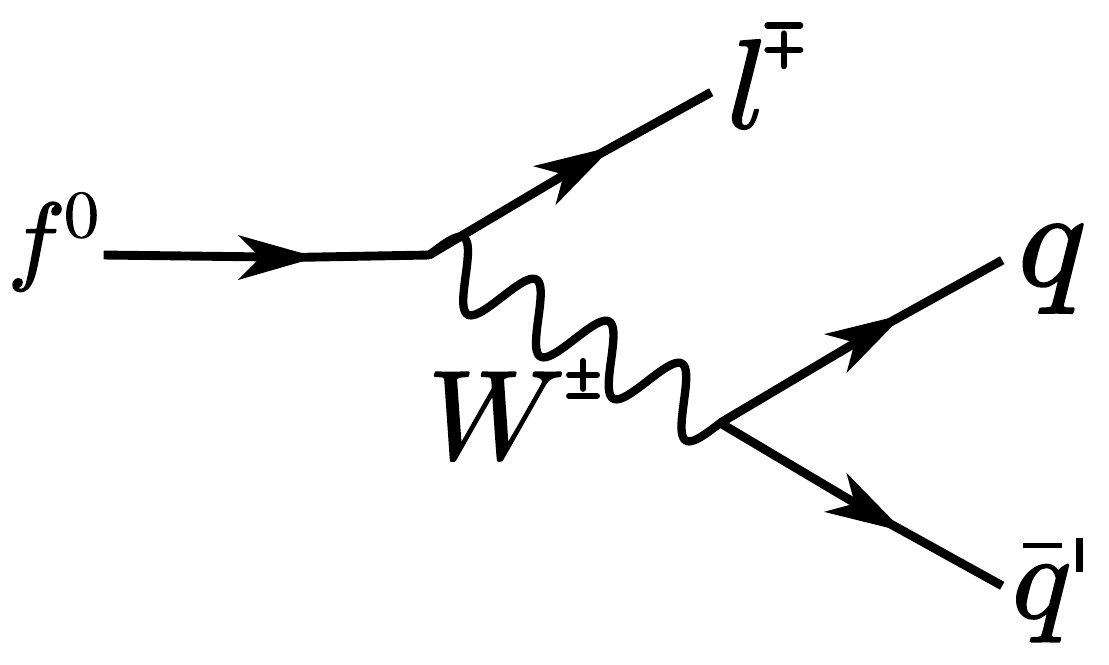}
  \caption{Example Feynman diagrams for the decays of exotic
    fermions appearing in our high-dimensional neutrino mass models.}
  \label{fig:dec}
\end{figure}

A number of different decay channels exist for the doubly charged
fermions, in principle. However, the most important decay for the
lightest doubly charged fermion is $f^{++}\to W^+ l_{\alpha}^+$, where
$\alpha=e,\mu,\tau$. The relative branching ratios to the different SM
lepton generations depends on the unknown Yukawa couplings. The total
final state from pair-produced $f^{++}f^{--}$ is then either two
opposite-sign charged leptons plus four jets, tri-lepton with missing
energy plus two jets or multilepton plus missing energy. From the
hadronic decays of the $W$ the mass of the doubly charged fermion can
be reconstructed. However, multilepton searches might lead to more
stringent lower limits on the exotic fermion mass, due to lower
backgrounds.  The CMS collaboration has recently published a search
based on multi-lepton final states \cite{Sirunyan:2017qkz}. From the
analysis presented in \cite{Cepedello:2017lyo} one can estimate that
this search implies a lower limits on the mass of $f^{++}$ of very
roughly [500,750] GeV. The range in this interval is due to the large
uncertainty in the branching ratio for the different final state
lepton generation. Given that the results of \cite{Sirunyan:2017qkz}
are based on $35.9/$fb, we expect that the full statistics of the
high-luminosity LHC will be sufficient to reach sensitivities up
to or in excess of $m \sim 2$ TeV. To obtain exact numbers would
require a MonteCarlo simulation of all backgrounds, which is beyond
the scope of this paper.

From the theoretical point of view, however, instead of deriving only
limits, it would be more interesting to establish lepton number
violation experimentally. For this, we have to turn to associated
production. Here, we have to consider the decays of $f^+$ and $f^0$,
see fig. (\ref{fig:prod}). $f^+$ can decay to either
$l^+_{\alpha}+Z^0$, $l^+_{\alpha}+h^0$ or $\nu+W^+$. Since we can not
determine experimentally the lepton number in events with missing
energy, we are not interested in the final state $\nu+W^+$, and only
the decays $Z^0 \to$ hadrons or $h \to b{\bar b}$ should be
considered when estimating the total number of events. $f^0$ can
decay to $l^{\pm}_{\alpha}+W^{\mp}$ and the decay to both charges of
leptons should occur with (nearly) equal branching ratios,
\footnote{At tree-level Br($f^+ \to l^{+}_{\alpha}+W^{-}$)= Br($f^+
  \to l^{-}_{\alpha}+W^{+}$), but for complex Yukawa couplings, small
  departures from equality are expected to occur at 1-loop level.}  as
indicated in fig. (\ref{fig:dec}). The total signal then consists of 
proton-proton collisions producing the final state $f^+f^0\to
l^+_{\alpha}l^{\pm}_{\beta}+4j$, which in the like-sign channel will
demonstrate the existence of lepton number violation. If we require 10
events before cuts in both the like-sign and opposite-sign dilepton
channels, optimistically we could expect to have sufficient sensitivity
at the HL-LHC for masses up to $m \sim 1.6$, $1.8$ and $2.0$ TeV for
the triplet, quadruplet and quintuplet fermions. Since this estimate
is based only on total cross sections and does not take into account
neither cuts nor backgrounds, it should be taken with a grain of salt.

Before closing this section, we want to briefly comment on other
fermions, not explicitly covered in this numerical calculation.  We
have chosen to discuss fermions, because all our models contain at
least one new fermion. In fact, $3_1^F$ appears in the majority of the
models that we have discussed. As fig. (\ref{fig:prod}) shows for pair
production, the most important quantity (apart from the fermion mass)
is the fermion charge. Thus, triply charged fermions, such as appear
in the multiplets ${\bf 4}_{3/2}^F$, ${\bf 6}_{1/2}^F$ and ${\bf 7}_{0}^F$ will have
{\em larger} cross sections and LHC searches should be able to
establish even stronger limits on models containing these particles.

Finally, we briefly comment on the exotic scalars. Scalars have
roughly a factor 4 smaller cross sections than fermions with the same
mass and quantum numbers. Generally one would thus expect the reach of
the LHC to be worse than for fermions. However, the large scalar
multiplets that appear in some of our constructions will have highly
distinctive final states. An example from pair produced scalars would
be 4 charged leptons with 4 $W$'s. Even larger multiplicities can
easily occur. Such states should have negligible standard model
backgrounds, thus partially compensating for the smaller cross
sections. It would be interesting to estimate the LHC reach more
quantitatively and we plan to do so in a future publication.

\section{Conclusions}
\label{sect:cncl}

We have discussed the systematic deconstruction of the $d=9$, $d=11$
and $d=13$ neutrino mass operators at tree-level. We have found all
genuine neutrino mass tree-level diagrams for these operators.  The
word ``genuine'' here refers to those diagrams which provide the
dominant contribution to the neutrino mass matrix, assuming no extra
symmetries beyond the standard model ones. Very few genuine models can
be constructed, despite the fact that the number of possible
topologies increases rapidly with the dimension of the operator: With
renormalizable vertices, one can build 18 topologies and 66 diagrams
at $d=9$ level; these numbers increase to 92 topologies and 504
diagrams at the $d=11$ level, and finally at $d=13$ one finds 576
topologies and 4199 diagrams. From all of these, we find only 10
genuine models: 2 models at $d=9$ and $d=11$ each, and 6 models at
$d=13$.

We have discussed how our definition of a genuine mass model requires
that all these high-dimensional models use large $SU(2)_L$
representations. For example, both of the two $d=9$ models require
quadruplets and quintuplets. On the other hand, for some $d=13$ models
scalar septets are needed.  These high-dimensional models require not
only larger representations but also more of them: Three new particles
are sufficient for one of the two $d=9$ models, while for $d=11$
($d=13$) already four (five) exotic fields are needed. Thus, models
become necessarily more baroque with larger dimensions.  This fact,
together with the rather low new physics scale required by the high
dimensionality of the operators, makes these models testable at
accelerator experiments and also in searches for lepton flavour
violation. We therefore plan to return to a study of the phenomenology
of these models in a future publication.

\begin{acknowledgements}
  We thank Ricardo Cepedello for pointing out to us a misidentified
  diagram in the first version of this draft.  This work was supported
  by the Spanish grants FPA2017-85216-P and SEV-2014-0398 (AEI/FEDER,
  UE), FPU15/03158 (MECD) and PROMETEOII/2018/165 (Generalitat
  Valenciana). J.C.H. is supported by Chile grants Fondecyt
  No. 1161463, Conicyt PIA/ACT 1406 and Basal FB0821. O.C-F. is
  supported by Chile grants Basal FB0821. R.F. also acknowledges the
  financial support from the Grant Agency of the Czech Republic,
  (GA\v{C}R), contract nr. 17-04902S, as well as from the grant Juan
  de la Cierva-formaci\'on FJCI-2014-21651 (from Spain).
\end{acknowledgements}

\appendix
\section{Neutrino mass and angle fits\label{sect:app}}

In the main text we gave simple estimates for the typical parameter
choices that generate a neutrino mass scale large enough to explain
the atmospheric neutrino oscillations. However, in all the models
presented in this paper it is actually easy to fit all angles and
masses simultaneously.  For the current status of oscillation data
see, for example, the recently updated global fit
\cite{deSalas:2017kay}.  In this appendix we briefly discuss how
tree-level neutrino mass models can be fitted to oscillation data.

We can divide all models discussed in the main text into just two
classes: (i) models in which only one type of exotic fermion couples
to the outside leptons, for example the $d=9$ model shown in
Fig.~\ref{fig:d9topos} on the left. And, (ii) models in which two
different fermions can couple to the leptons, for example the $d=9$
model in Fig.~\ref{fig:d9topos} on the right.

We start with case (i). First, recall that neutrino oscillations
require at least two neutrino masses to be non-zero. For models of
case (i) there will be one non-zero neutrino mass for each copy of
exotic fermions coupling to the leptons. Assuming there are three
copies of these exotic fermions one can then use a slight modification
of the well-known Casas-Ibarra parametrization~\cite{Casas:2001sr}
that makes it possible to fit neutrino data for an ordinary seesaw ($d=5$
tree-level). In the simplest seesaw, the light neutrino mass matrix is
approximately given by
\begin{equation}
  \label{eq:seesaw}
  m_{\nu} = - m_D^T ({\hat M}_R)^{-1}m_D,
\end{equation}
where $m_D$ is the Dirac mass term for neutrinos and ${\hat M}_R$ is
the diagonal matrix of the heavy neutrino eigenvalues. Diagonalizing
the light neutrino mass matrix with a matrix $V_L$ and solving
Eq.~\eqref{eq:seesaw} for $m_D$ one finds~\cite{Casas:2001sr}
\begin{equation}
  \label{eq:CI}
  m_D = i \sqrt{{\hat M}_R} {\cal R} \sqrt{{\hat m}_{\nu}} V_L^{\dagger}
\end{equation}
${\cal R}$ is a matrix of three complex angles, with ${\cal R}^T{\cal
  R}=1$, left undetermined when solving Eq.~\eqref{eq:seesaw}.  $V_L$
contains the measured neutrino angles and Dirac CP-phase $\delta$ and
${\hat m}_{\nu}$ is the diagonal matrix of the light neutrino eigenvalues.

The derivation of Eq.~\eqref{eq:CI} relies on the fact that in the
standard model augmented with a simple seesaw one can always perform a
basis change, such that $M_R$, the mass matrix for the right-handed
neutrinos, is diagonal. In the higher dimensional neutrino mass
models, discussed in this paper, for the effective neutrino mass, see
Eq.~\eqref{eq:seesaw}, we have to replace $M_R$ by a product of
matrices.  For example for the $d=9$ model one finds:
\begin{equation}
  \label{eq:mRd9}
  M_R^{-1} \to M_{eff}^{-1}
  ={\hat M}_3^{-1} m_{34}^T {\hat M}_4^{-1} m_{45}^T {\hat M}_5^{-1} m_{45}
  {\hat M}_4^{-1} m_{34} {\hat M}_3^{-1} .
\end{equation}
Basis changes can be used to diagonalize the
vector-like mass terms---but not the ``Dirac''-like mass terms
$m_{ij}$ at the same time. $M_{eff}$ is a complex symmetric matrix
and can be diagonalized with a matrix $U$, containing in general
3 angles and 3 phases. For arbitrary choices of the parameters entering
the various matrices in Eq.~\eqref{eq:mRd9} one can find $U$ numerically
and then use Eq.~\eqref{eq:CI} to determine the correct choice
of $m_D$, using the simple replacement:
\begin{equation}\label{eq:ModCI}
  \sqrt{{\hat M}_R} \to U^{\dagger}\sqrt{{\hat M}_{eff}}.
\end{equation}
For other models of the same type the form of $M_{eff}$ may change,
but the procedure for the neutrino fit is completely analogous. 

For case (ii) one can do a neutrino fit using only one copy of each of
the two exotic fermions coupling to leptons. Let ${\vec h^1}$ and
${\vec h^2}$ be the two Yukawa vectors coupling exotic fermions to
standard model leptons in any given model of this type. Then,
schematically, one finds a neutrino mass matrix given by:
\begin{equation}
  \label{eq:mnu2}
  (m_{\nu})_{\alpha\beta} = c (h^1_{\alpha}h^2_{\beta} + h^1_{\beta}h^2_{\alpha})
\end{equation}
Here, $c$ is a constant with dimension of mass. For example, in the
$d=9$ model shown on the right of Fig.~\ref{fig:d9diags}, $c$ is given
by $c=Y_{45}\lambda_4 \frac{\mu_3v_{SM}^6} {M_{\bf 5_1} M_{\bf 4}
m_4^2m_3^2}$. The matrix in Eq.~\eqref{eq:mnu2} has determinant
zero. Thus, it can solved analytically using only quadratic equations.
Let $|{\vec h^i}|$ be the absolute value of the vector ${\vec h^i}$.
Then the two non-zero of eigenvalues Eq.~\eqref{eq:mnu2} are given by:
\begin{equation}
  \label{eq:fit2}
  m_{\nu_{2,3}} = c \Big(
  {\vec h^1}\cdot{\vec h^2}\mp |{\vec h^1}||{\vec h^2}|\Big). 
\end{equation}
Neutrino angles, on the other hand, depend only on ratios of entries
in the Yukawa vectors. Although also the eigenvectors can be found
analytically, neutrino angles are fitted most easily numerically. We
calculate $m_{\nu}$ from the measured $\Delta m^2_{ij}$ and
$\theta_{ij}$.  Then, for any choice of the parameters entering $c$,
we can choose one entry in the two Yukawa vectors freely, say for
example $h^1_1$, and numerically solve five entries of the matrix in
Eq.~\eqref{eq:mnu2} for five independent entries in $m_{\nu}$. Note
that, since $c$ itself can contain small parameters (for example, all
of $Y_{45}$, $\lambda_4$ and $\mu_3$ can be small), one needs to check
that the resulting Yukawa vectors have entries which are perturbative.

We close this short appendix with a comment. In certain limits, the
two types of models can be fitted with both procedures described
above.  For example, a model in class (ii) could have 3 copies of both
exotic fermions. If ${\vec h^1} \propto {\vec h^2}$ for the three
pairs of vectors, one can also use the modified Casa-Ibarra procedure
to find solutions fitting all data.

\section{RGEs and $\beta$-coefficients for the different
  models\label{subsect:RGE}}

Ignoring the minor contribution from Yukawa couplings, the
renormalization group equation for the gauge couplings can be written
at 2-loop order as:
\begin{equation}\label{eq:rge}
  \frac{dg_i}{dt} = b_i \frac{g_i^3}{(4\pi)^2} + B_{ij}\frac{g_i^3g_j^2}{(4 \pi)^4}
\end{equation}

For the Standard Model, using the ordering $U(1)_Y$,$SU(2)_L$ and $SU(3)_c$, 
the coefficients are the following:
\begin{equation}
  b=\left(\frac{41}{10},-\frac{19}{6},-7\right)\textrm{ and }
  B=\left(\begin{array}{ccc}
\frac{199}{50} & \frac{27}{10} & \frac{44}{5}\\
\frac{9}{10} & \frac{35}{6} & 12\\
\frac{11}{10} & \frac{9}{2} & -26
\end{array}\right)\,.
\end{equation}

In section (\ref{subsubsect:RGE}) we have shown numerical results
for the running of $g_2$ for six different model variants, two
models each at $d=9$, $d=11$ and $d=13$. Here, for convenience,
we repeat the particle content of those six models and give their
1- and 2-loop renormalization group running coefficients (note that those involving $SU(3)_c$,  i.e. $b_3$, $B_{3i}$ and $B_{i3}$, are the same as in the SM). We considered the cases where the number of copies  $n$ of each new fermion representation is either 1 or 3.

\bigskip
\noindent{\bf Model-I, $d=9$ (new fields: $\mathbf{3}_{\pm1}^{F}$, $\mathbf{4}_{\pm1/2}^{F}$,
	$\mathbf{5}_{0}^{F}$)}
~
\begin{align}
b=\left(\frac{73}{10},\frac{77}{6},-7\right)\textrm{ and }B= & \left(\begin{array}{ccc}
\frac{433}{50} & \frac{261}{10} & \frac{44}{5}\\
\frac{87}{10} & \frac{2261}{6} & 12\\
\frac{11}{10} & \frac{9}{2} & -26
\end{array}\right)\quad\quad\left[n=1\right]\\
b=\left(\frac{137}{10},\frac{269}{6},-7\right)\textrm{ and }B= & \left(\begin{array}{ccc}
\frac{901}{50} & \frac{729}{10} & \frac{44}{5}\\
\frac{243}{10} & \frac{6713}{6} & 12\\
\frac{11}{10} & \frac{9}{2} & -26
\end{array}\right)\quad\quad\left[n=3\right]
\end{align}

\noindent{\bf Model-II, $d=9$ (new fields: $\mathbf{4}_{3/2}^{S}$, $\mathbf{5}_{\pm1}^{F}$,
	$\mathbf{4}_{\pm1/2}^{F}$, real $\mathbf{3}_{0}^{S}$)}
~
\begin{align}
b=\left(\frac{107}{10},\frac{113}{6},-7\right)\textrm{ and }B= & \left(\begin{array}{ccc}
\frac{407}{10} & \frac{1647}{10} & \frac{44}{5}\\
\frac{549}{10} & \frac{3671}{6} & 12\\
\frac{11}{10} & \frac{9}{2} & -26
\end{array}\right)\quad\quad\left[n=1\right]\\
b=\left(\frac{203}{10},\frac{353}{6},-7\right)\textrm{ and }B= & \left(\begin{array}{ccc}
\frac{2791}{50} & \frac{3267}{10} & \frac{44}{5}\\
\frac{1089}{10} & \frac{9851}{6} & 12\\
\frac{11}{10} & \frac{9}{2} & -26
\end{array}\right)\quad\quad\left[n=3\right]
\end{align}

\noindent{\bf Model-I, $d=11$ (new fields: $\mathbf{3}_{\pm1}^{F}$, $\mathbf{5}_{2}^{S}$,
	$\mathbf{5}_{1}^{S}$, real $\mathbf{3}_{0}^{S}$)}
~
\begin{align}
b=\left(\frac{23}{2},\frac{13}{2},-7\right)\textrm{ and }B= & \left(\begin{array}{ccc}
\frac{1307}{10} & \frac{3771}{10} & \frac{44}{5}\\
\frac{1257}{10} & \frac{1129}{2} & 12\\
\frac{11}{10} & \frac{9}{2} & -26
\end{array}\right)\quad\quad\left[n=1\right]\\
b=\left(\frac{163}{10},\frac{71}{6},-7\right)\textrm{ and }B= & \left(\begin{array}{ccc}
\frac{6967}{50} & \frac{4059}{10} & \frac{44}{5}\\
\frac{1353}{10} & \frac{3899}{6} & 12\\
\frac{11}{10} & \frac{9}{2} & -26
\end{array}\right)\quad\quad\left[n=3\right]
\end{align}

\noindent{\bf Model-II, $d=11$ (new fields: $\mathbf{3}_{\pm1}^{F}$, $\mathbf{5}_{2}^{S}$,
	$\mathbf{6}_{3/2}^{S}$, $\mathbf{5}_{1}^{S}$, real $\mathbf{3}_{0}^{S}$)}
~
\begin{align}
b=\left(\frac{71}{5},\frac{37}{3},-7\right)\textrm{ and }B= & \left(\begin{array}{ccc}
\frac{4361}{25} & \frac{3303}{5} & \frac{44}{5}\\
\frac{1101}{5} & \frac{3601}{3} & 12\\
\frac{11}{10} & \frac{9}{2} & -26
\end{array}\right)\quad\quad\left[n=1\right]\\
b=\left(19,\frac{53}{3},-7\right)\textrm{ and }B= & \left(\begin{array}{ccc}
\frac{4577}{25} & \frac{3447}{5} & \frac{44}{5}\\
\frac{1149}{5} & \frac{3857}{3} & 12\\
\frac{11}{10} & \frac{9}{2} & -26
\end{array}\right)\quad\quad\left[n=3\right]
\end{align}

\noindent{\bf Model-I, $d=13$ (new fields: $\mathbf{3}_{\pm1}^{F}$, $\mathbf{4}_{\pm3/2}^{F}$,
	$\mathbf{5}_{\pm1}^{F}$, $\mathbf{6}_{\pm1/2}^{F}$, $\mathbf{5}_{0}^{F}$)}
~
\begin{align}
b=\left(\frac{189}{10},\frac{99}{2},-7\right)\textrm{ and }B= & \left(\begin{array}{ccc}
\frac{226}{5} & \frac{1008}{5} & \frac{44}{5}\\
\frac{336}{5} & 1596 & 12\\
\frac{11}{10} & \frac{9}{2} & -26
\end{array}\right)\quad\quad\left[n=1\right]\\
b=\left(\frac{97}{2},\frac{929}{6},-7\right)\textrm{ and }B= & \left(\begin{array}{ccc}
\frac{3191}{25} & \frac{2997}{5} & \frac{44}{5}\\
\frac{999}{5} & \frac{14329}{3} & 12\\
\frac{11}{10} & \frac{9}{2} & -26
\end{array}\right)\quad\quad\left[n=3\right]
\end{align}

\noindent{\bf Model-II, $d=13$ (new fields: $\mathbf{4}_{3/2}^{S}$, $\mathbf{5}_{\pm1}^{F}$,
	$\mathbf{6}_{\pm1/2}^{F}$, $\mathbf{5}_{0}^{F}$)}

\begin{align}
b=\left(\frac{111}{10},\frac{251}{6},-7\right) & \textrm{ and }B=\left(\begin{array}{ccc}
\frac{1022}{25} & \frac{936}{5} & \frac{44}{5}\\
\frac{312}{5} & \frac{4480}{3} & 12\\
\frac{11}{10} & \frac{9}{2} & -26
\end{array}\right)\quad\quad\left[n=1\right]\\
b=\left(\frac{43}{2},\frac{257}{2},-7\right) & \textrm{ and }B=\left(\begin{array}{ccc}
\frac{1409}{25} & \frac{1971}{5} & \frac{44}{5}\\
\frac{657}{5} & 4305 & 12\\
\frac{11}{10} & \frac{9}{2} & -26
\end{array}\right)\quad\quad\left[n=3\right]
\end{align}



\end{document}